\documentclass[preprint2]{aastex63}

\usepackage{xspace}
\usepackage{amsfonts}
\usepackage[figuresright]{rotating}
\newcommand{\tclean}{\texttt{tclean}\xspace}
\newcommand{\nupper}{\ensuremath{N_u}\xspace}

\newcommand{\methanol}{CH$_3$OH\xspace}
\newcommand{\sgrbtwosouth}{Sgr B2(S)\xspace}
\newcommand{\dsi}{DS1\xspace}
\newcommand{\dsii}{DS2\xspace}
\newcommand{\dsiii}{DS3\xspace}
\newcommand{\dsiv}{DS4\xspace}
\newcommand{\dsv}{DS5\xspace}
\newcommand{\dsvi}{DS6\xspace}
\newcommand{\dsvii}{DS7\xspace}
\newcommand{\dsviii}{DS8\xspace}
\newcommand{\dsix}{DS9\xspace}
\newcommand{\formaldehyde}{H$_2$CO\xspace}
\newcommand{\deepsouth}{Sgr B2(DS)\xspace}

\submitjournal{ApJ}

\shorttitle{Thermal Properties of Sgr B2(DS) Hot Cores}
\shortauthors{Jeff et al.}
\graphicspath{{./}{figures/}}

\begin{document}

\title{Thermal Properties of the Hot Core Population in Sagittarius B2 Deep South}

\author[0000-0003-0416-4830]{Desmond Jeff}
\correspondingauthor{Desmond Jeff}
\affil{Department of Astronomy, University of Florida \\
211 Bryant Space Science Center
P.O Box 112055
Gainesville, FL 32611-2055 USA}
\affil{National Radio Astronomy Observatory, 520 Edgemont Road, Charlottesville, VA 22903-2475, USA}
\email{d.jeff@ufl.edu}

\author[0000-0001-6431-9633]{Adam Ginsburg}
\affil{Department of Astronomy, University of Florida \\
211 Bryant Space Science Center
P.O Box 112055
Gainesville, FL 32611-2055 USA}

\author[0000-0002-4407-885X]{Alyssa Bulatek}
\affil{Department of Astronomy, University of Florida \\
211 Bryant Space Science Center
P.O Box 112055
Gainesville, FL 32611-2055 USA}

\author[0000-0002-0533-8575]{Nazar Budaiev}
\affil{Department of Astronomy, University of Florida \\
211 Bryant Space Science Center
P.O Box 112055
Gainesville, FL 32611-2055 USA}

\author[0000-0002-3078-9482]{\'Alvaro S\'anchez-Monge}
\affil{Institut de Ci\`encies de l'Espai (ICE, CSIC), Carrer de Can Magrans s/n, E-08193, Bellaterra, Barcelona, Spain}
\affil{Institut d'Estudis Espacials de Catalunya (IEEC), Barcelona, Spain}

\author[0000-0001-6551-6444]{M\'elisse Bonfand}
\affil{Departments of Astronomy \& Chemistry, University of Virginia, Charlottesville, VA 22904, USA}

\author[0000-0002-6073-9320]{Cara Battersby}
\affil{University of Connecticut, Department of Physics \\
196A Auditorium Road Unit 3046 
Storrs, CT 06269 USA}
\author[0000-0002-5927-2049]{Fanyi Meng}
\affil{University of Chinese Academy of Sciences, Beijing 100049, China}
\author[0000-0003-2141-5689]{Peter Schilke}
\affil{I. Physikalisches Institut, Universit\"at zu K\"oln \\ 
Z\"ulpicher Str. 77, 50937 K\"oln, Germany}
\author[0000-0002-1730-8832]{Anika Schmiedeke}
\affil{Green Bank Observatory, 155 Observatory Rd P.O. Box 2. Green Bank, WV 24944, USA}

\begin{abstract}
We report the discovery of 9 new hot molecular cores in the Deep South (DS) region of Sagittarius B2 using Atacama Large Millimeter/submillimeter Array Band 6 observations. We measure the rotational temperature of \methanol and derive the physical conditions present within these cores and the hot core \sgrbtwosouth. The cores show heterogeneous temperature structure, with peak temperatures between 252 and 662 K. We find that the cores span a range of masses (203-4842 M$_\odot$) and radii (3587--9436 AU). \methanol abundances consistently increase with temperature across the sample. Our measurements show the DS hot cores are structurally similar to Galactic Disk hot cores, with radii and temperature gradients that are comparable to sources in the Disk. They also show shallower density gradients than Disk hot cores, which may arise from the Central Molecular Zone’s higher density threshold for star formation. The hot cores have properties which are consistent with those of Sgr B2(N), with 3 associated with Class II \methanol masers and one associated with an UCHII region. Our sample nearly doubles the high-mass star forming gas mass near \sgrbtwosouth and suggest the region may be a younger, comparably massive counterpart to Sgr B2(N) and (M). The relationship between peak \methanol abundance and rotational temperature traced by our sample and a selection of comparable hot cores is qualitatively consistent with predictions from chemical modeling. However, we observe constant peak abundances at higher temperatures ($T\gtrsim250$ K), which may indicate mechanisms for methanol survival that are not yet accounted for in models.

\end{abstract}

%\keywords{empty}

\section{Introduction} \label{sec:intro}
The Central Molecular Zone (CMZ) is the closest ``extreme" environment in which we can resolve individual sites of star formation. It is characterized by high molecular gas densities ($\sim10^4$ cm$^{-3}$, \citet{Lis&Carlstrom1994, Mills+2018}) and enrichment with complex molecules \citep{Belloche2013}. This environment is analogous to environments that were present near the peak in the cosmic star formation rate (z$\sim2$, \citet{Kruijssen&Longmore2013}), providing a vital laboratory within which to investigate how star formation proceeded during that era and in current sites of vigorous star formation (e.g., LIRGs and ULIRGs, \citet{Armus2009}). Massive stars are a crucial component in these investigations, given the role they play in injecting various kinds of feedback into the interstellar medium of their host galaxies (e.g., \citet{Meurer+1995,Zinnecker&Yorke2007,Ramachandran+2018}). %However, its star formation rate is at least an order of magnitude lower than predictions made with surface density or volumetric relations \citep{Longmore2013,Barnes2017}. This poses a challenge to these relations as, in spite of their broad applicability in both local ($\sim$500 pc) and extragalactic environments, they are unable to accurately predict the star formation rate in the CMZ's unique environment. Among others, \citet{Barnes2017} cite a bottom-heavy initial mass function (IMF) as a possible solution to this discrepancy, and there is observational evidence supporting the existence of such IMFs in early-type galaxies (e.g. \citet{Cappellari2012}). However, studies of the IMF in the CMZ have yielded evidence of the opposite scenario: young massive clusters within the Galactic Center show evidence of a top-heavy or bottom-light IMF \citep{Bartko2010, Lu+2013, Hosek2019}. This has spurred intense investigation into the physical conditions within the CMZ and how mechanisms like feedback from massive young stellar objects (YSOs) and turbulence may alter the IMF (e.g. \citet{Ginsburg2016, Walker2021}.
However, owing to their distance and brevity of their lifetimes, the earliest phases of massive stellar evolution are difficult to observe (e.g., \citet{Zinnecker&Yorke2007,Battersby+2017, Motte+2018}). Massive protostars reach the main sequence while still heavily embedded in their natal envelopes, preventing direct observation of the earliest phases of their formation processes. Instead, we can study the composition and physical conditions within the protostellar envelope to infer the characteristics of the embedded protostar. 

Located $\sim100$ pc in projection away from Sgr A$^*$, the CMZ molecular cloud Sagittarius B2 (Sgr B2) is the most massive ($\sim10^7$ M$_\odot$) molecular cloud in the galaxy \citep{Lis&Goldsmith1990}. It shows a high degree of chemical enrichment, with many molecules being detected first in Sgr B2 \citep{McGuire2018} and many more continuing to be found \citep{Belloche2018}. It has also been described as undergoing a``mini-starburst" event, containing $>70$ high-mass stars, Class II methanol (\methanol) masers, water (H$_2$O) masers, diffuse and ultra-compact HII (UCHII) regions, and $>200$ 3 mm compact continuum sources that have been identified as YSOs (e.g., \cite{Ginsburg2018}, \cite{Budaiev+2024}). Several of these 3 mm continuum sources have been further identified as hot molecular cores \citep{Bonfand2017,SanchezMonge+2017}, which are sites of high-mass star formation that show emission from rare and complex molecular species \citep{Cesaroni2005}. These sources are thought to represent a key phase in massive protostellar evolution and the chemical evolution of the ISM, as the antecedents to UCHII regions and formation sites of complex organic molecules (COMs, \citet{Herbst&vanDishoeck2009}). The presence of hot cores in Sgr B2 thus makes it an ideal region within which to observe ongoing high-mass star formation and its influence on the physical and chemical properties of the surrounding gas. %It is also ideal for observing the so-called ``pre-IMF" or core mass function (CMF), as many of the stars throughout the cloud remain embedded in their natal envelopes.

Sgr B2 can be subdivided into four main regions: the protocluster complexes, Sgr B2(N)orth, (M)ain, and (S)outh; and the molecular ridge, Deep South (DS) \citep{Schmiedeke2016}. Prior studies in this region have primarily focused on Sgr B2(N) and (M) due to their relative brightness, chemical enrichment, and numerous sites of active high mass star formation. This has left \sgrbtwosouth and (DS) relatively unexplored until recently, with the discovery of an arc of YSOs and non-thermal radio emission in (DS) \citep{Ginsburg2018, Meng2019}. Given that (DS) only contains one known HII region in \sgrbtwosouth \citep{Lu+2019,Meng+2022} and numerous embedded YSOs, it is thought that it is in an earlier phase of star formation than its counterparts in (N) and (M). It therefore represents a unique opportunity to observe the early phases of high mass star formation in the CMZ.

\methanol is a molecule of particular interest in star formation studies. It is the simplest COM and is believed to be one of several ``parent" molecules necessary for the formation of more complex COMs \citep{Nomura&Millar2004,Garrod&Herbst2006,Taquet+2016}. %observed to have high abundances toward CMZ clouds \citep{Requena-Torres+2008}.
The abundances of COMs are often measured against the abundance of \methanol when testing the effects of physical conditions on COM formation rates \citep{Requena-Torres+2008,Herbst&vanDishoeck2009,Bonfand2019}. However, in spite of its utility and importance in star formation and astrochemical studies, the conditions leading to \methanol formation at the observed abundances remain under investigation \citep{SanchezMonge+2017,Bonfand2019}. This is especially true in the CMZ.

In this paper, we report the discovery and the \methanol-derived physical characteristics of 9 hot molecular cores through the Deep South (DS) region of Sgr B2. In Section \ref{sec:methods}, we describe our data-reduction and analysis procedure. In Section \ref{sec:results}, we discuss the spectral characteristics of the hot cores and their thermochemical properties. In Section \ref{sec:discussion}, we discuss the biases that may affect our \methanol temperature and abundance measurements, physical mechanisms which may be responsible for our measurements, and compare the hot cores of \deepsouth to a selection of Galactic Disk hot cores and neighboring hot cores in Sgr B2(N). We summarize our conclusions in Section \ref{sec:conclusion}.

\section{Methods} \label{sec:methods}
 
 \subsection{Data reduction and imaging}
\begin{figure*}
    \centering
    \includegraphics[scale=0.6]{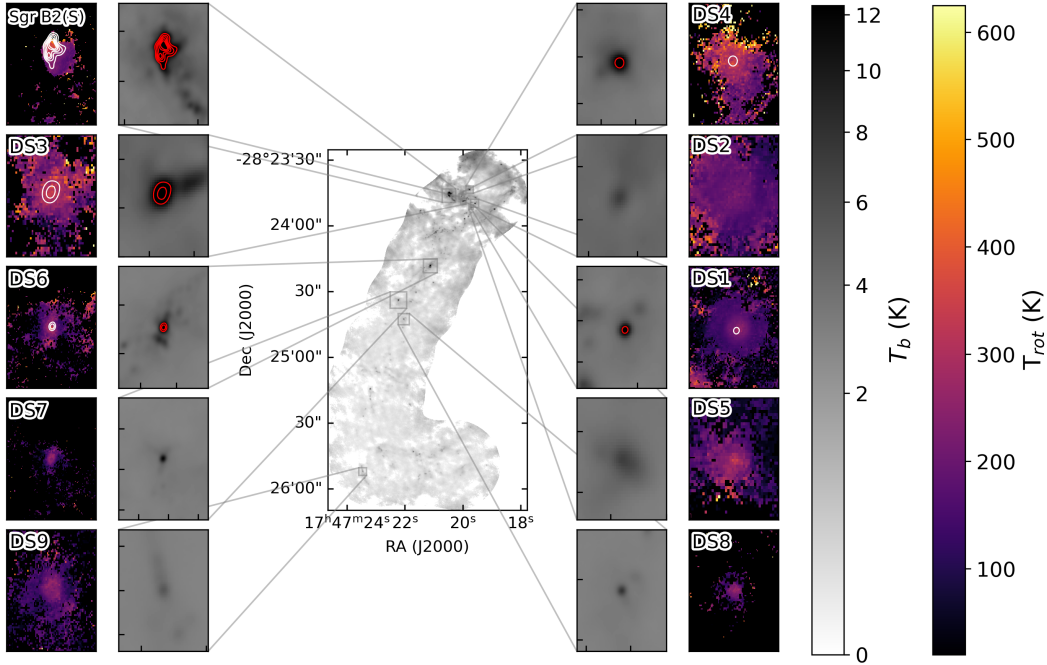}
    \caption{1 mm dust continuum of Sgr B2(DS) after feathering with Bolocam Galactic Survey data of Sgr B2 \citep{Ginsburg+2013}. The inset images show the \methanol temperature maps from each hot core (outer) and its corresponding continuum emission (inner). Red contours indicate regions where the continuum brightness exceeds 12 K (levels are at 12, 18, 24, 30, and 36 K).}
    \label{fig:zoomfig}
\end{figure*}

We have carried out this work using 12 m Band 6 data collected as a part of ALMA Project Code 2017.1.00114.S (PI: Ginsburg). The full dataset consists of 10 fields that cover the arc of 3 mm continuum sources identified in \citet{Ginsburg2018}, as shown in Figure \ref{fig:zoomfig}.

Each field contains four spectral windows, centered on 217.35, 219.20, 231.30, and 233.18 GHz, respectively. Each spectral window has a bandwidth of 1.875 GHz and a spectral resolution of 488 kHz ($\sim0.6$ km s$^{-1}$). Data cubes for each spectral window were produced using a robust parameter of 0 and by cleaning to a threshold of 3$\sigma$ above the noise (3.09 mJy) using \tclean (CASA version 5.6.0, \citet{McMullin2007}).

Due to the density of emission lines in chemically rich sources like hot cores, identifying line-free channels for continuum subtraction is challenging. After identifying sources (See Section \ref{subsec:sourceid}), we use the \texttt{--continuum} mode of STATCONT \citep{SanchezMonge2018} to perform continuum subtraction on each spectral window. This allows us to identify the hot core's continuum level and associated standard deviation ($\sigma$) in each pixel. 

We produced a continuum image by combining all four spectral windows with \texttt{tclean}. It was cleaned to a threshold of 1 mJy and imaged with a Briggs robust parameter of 2, achieving a beam size of $0.36\arcsec\times 0.29\arcsec$. The largest angular scale (LAS) of our observations was 2.075'', and our pixel scale was 0.05''. To account for filtering of large-scale continuum emission, we use \texttt{uvcombine} to feather the Atacama Large Millimeter/submillimeter Array (ALMA) observations with 1.1 mm continuum data from the Bolocam Galactic Plane Survey \citep{Ginsburg+2013}. The LAS in their work is $\sim2\arcmin$. The lowest noise level away from bright sources in the feathered continuum image is $\sim0.2$ mJy.

\subsection{Data Analysis}
\subsubsection{Source Identification}\label{subsec:sourceid}
To identify hot core candidates, we begin by performing a by-eye examination of the 1 mm molecular line cubes. Canonical hot cores are compact sources, with radii $\lesssim0.1$ pc (e.g., \citep{Cesaroni2005}). We therefore define candidate hot cores as sources that have significantly detected ($>3\sigma$), centrally-peaked, and compact ($<0.1$ pc, 2'' at our resolution) \methanol emission. To identify \methanol lines in the sources, we use \texttt{astroquery} to search the JPL Millimeter, Submillimeter, and Microwave Spectral Line Catalog \citep{PICKETT1998883} for lines within the given spectral window using Splatalogue (\texttt{astroquery}, \citet{Ginsburg2016astroquery}; Splatalogue, \citet{MarkwickKemper2006}). Given its brightness, we examine the first spectral window for the $5_1-4_2$ vt=0 transition of \methanol. Once identified, we use the doppler velocity of this line as a tentative velocity for the source. We then fit a single-component Gaussian to the line using the Levenberg-Marquardt least-squares fitter in \texttt{astropy} to obtain a precise doppler velocity for the $5_1-4_2$ line based on its primary velocity component.

Emission from low-J ($E_{upper}<96$ K) \methanol lines traces both compact sources and extended emission. As such, we choose the detection of the $10_{2-}-9_{3-}$ line to mark hot core candidates ($E_{upper} = 165$ K, see Table \ref{tab:detectedlines}), as it is the lowest-J transition that uniquely traces compact sources. We additionally exclude sources that are beyond the half-power beam width (HPBW) of our observations to minimize the effects of the primary-beam correction on our measurements. Temperature measurements performed on sources which are within and outside the HPBW in different fields show temperatures that are discrepant by factors of a few to several.

Once a candidate has been identified in the line data, we cross-correlate the position of the \methanol source to the 1 mm continuum data. We create square cutout images centered on the corresponding continuum peak that enclose the source and the surrounding 2--6.25'' ($0.08-0.25$ pc). This prevents the inclusion of additional sources with different doppler velocity than the selected source, which would result in spurious temperature values after the following analysis steps are complete. For the source \sgrbtwosouth, which is comprised of multiple continuum sources sharing a molecular envelope, we center on the reference pixel from which we pulled spectra (see Table \ref{table:obsprops}).

\begin{deluxetable}{ccc}\label{tab:detectedlines}

\tablecaption{Detected \methanol Lines}

\tablehead{\colhead{Transition} & \colhead{Rest Frequency} & \colhead{$E_U$}\\ 
\colhead{} & \colhead{(GHz)} & \colhead{(K)}}

\startdata
$5_{1}-4_{2}$ E1 & 216.946 & 55 \\
$6_{1-}-7_{2-}$ $v_t$=1 & 217.299 & 373 \\
$15_{6\pm}-16_{5\pm}$ $v_t$=1 & 217.643 & 745 \\
$20_{1}-20_{0}$ E1 & 217.887 & 508 \\
$4_{2}-3_{1}$ E1 & 218.440 & 45 \\
$25_{3}-24_{4}$ E1 & 219.984 & 802 \\
$23_{5}-22_{6}$ E1 & 219.994 & 775 \\
$8_{0}-7_{1}$ E1 & 220.079 & 96 \\
$10_{2-}-9_{3-}$ & 231.281 & 165 \\
$10_{2+}-9_{3+}$ & 232.419 & 165 \\
$8_{6}-8_{7}$ E1 $v_t$=1 & 232.645 & 675 \\
$18_{3+}-17_{4+}$ & 232.784 & 446 \\
$14_{6}-14_{7}$ E1 $v_t$=1 & 232.925 & 835 \\
$10_{-3}-11_{-2}$ E2 & 232.946 & 190 \\
$18_{3-}-17_{4-}$ & 233.796 & 446 \\
$13_{3-}-14_{4-}$ $v_t$=2 & 233.917 & 868 \\
$13_{3+}-14_{4+}$ $v_t$=2 & 233.917 & 868 \\
\enddata

\end{deluxetable}

\subsection{Measuring \methanol rotational temperature} \label{subsec:trotmeasurement}
As hot cores are chemically enriched and often show broad or complex velocity structure, there may be contamination by non-\methanol species that must be accounted for before beginning analysis. To mitigate the inclusion of contaminants, we examine the profiles of the \methanol lines in the source to identify a bright, isolated line from which to derive a representative velocity field for the gas. Lines that are not fully blended (i.e., contaminants $\gtrsim$ 1$\sigma$ line width separated from the chosen line) may also be used after limiting the range of included velocities to the primary emission component. We chose a different isolated reference line for each hot core. Frequently chosen transitions include $5_1-4_2$ vt=0, $8_0-7_1$ vt=0, and $20_1-20_0$ vt=0 (see Table \ref{table:obsprops}). Once a reference line has been selected, we create spectral slabs centered on the measured velocity with widths equal to the FWHM of the representative line. We then apply a 3$\sigma_{SC}$ signal mask to the slab, where $\sigma_{SC}$ is the uncertainty on the continuum level derived from STATCONT. and compute the velocity field (moment-1) and FWHM (moment-2) maps of the core for use in forthcoming analysis steps.

We search for detections of methanol lines by creating LTE models at temperatures and column densities corresponding to our first rough estimates, i.e., $T=150-300$ K and $N_{tot} = 0.1-1.0 \times10^{17}$ cm$^{-2}$.
We then compare the modeled intensity to the measured uncertainty at each frequency and select those transitions with predicted $>3\sigma$ detections.
We then ignore other lines of \methanol, such as those with high upper-state energy levels (several have $E_U\gtrsim2000$ K).
Once a candidate line is identified, we check whether the data within the selected velocity range ($v_{center}\pm FWHM$) have a peak intensity $T_{b,max} \geq3\sigma_{SC}$.
We calculate the integrated intensity of lines that meet this criterion and exclude others in the remainder of our modeling. The velocity ranges vary between sources, and we have indicated as such in the caption of each figure in the associated figure set (see Appendix \ref{app:moment0s}).

\begin{figure}
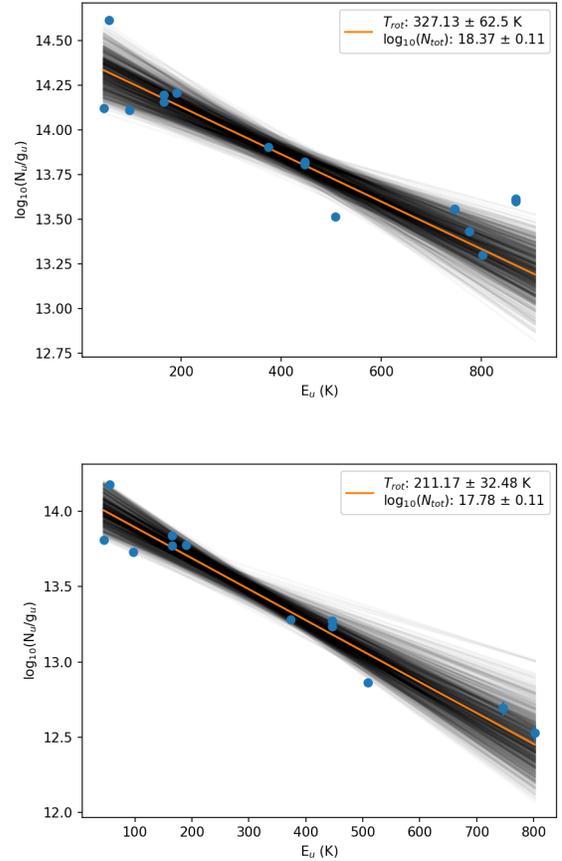

    \centering
    \includegraphics[scale=0.5]{f2.pdf}
    \includegraphics[scale=0.5]{f3.pdf}
    \caption{Rotational diagrams representative of the hot core envelope in {\sgrbtwosouth (upper)} and over the hot core continuum peak from  and \dsii (lower). The cores show \nupper values (blue points) that are reasonably well fit by the LTE model (orange line), with some scatter. The black background lines show the LTE models derived from the bootstrapping method.}%The values of \nupper are well fit by model gas that is in LTE, and this behavior is also seen in gas of similar temperatures ($\sim 160$ K) in \dsi. (lower) Rotational diagram from gas near the temperature peak in \dsi. This gas shows complex \nupper behavior that is consistent with non-LTE excitation and contamination from other molecular species. }
    \label{fig:rotdiags}
\end{figure}

The spectra we analyze are continuum-subtracted. However, the continuum brightnesses in the hot cores can contribute between 20\% to 40\% of the line brightnesses, which dilutes the observed integrated intensity if not accounted for. We therefore apply a final correction for continuum dilution by adding the continuum brightness multiplied by the measured FWHM of each line to the integrated intensity. Using these corrected integrated intensity maps (see Appendix \ref{app:moment0s}), we calculate upper-state column densities of the lines using the following equation:
\begin{equation}
    N_u=\frac{8\pi k_B\nu^2}{hc^3A_{ij}}\int_{v_1}^{v_2}T_bdv
    \label{eq:nupper}
\end{equation} 
where $N_u$ is the upper-state column density, $k_B$ is the Boltzmann constant, $\nu$ is the rest frequency of the transition, $h$ is the Planck constant, $c$ is the speed of light, $A_{ij}$ is the Einstein $A$ coefficient for the transition, and $\int_{v_1}^{v_2}T_bdv$ is the integrated intensity \citep{MangumandShirley2015}. By dividing $N_u$ by $g_u$, the degeneracy of the transition, and plotting with respect to the upper-state column density $E_u$, we construct per-pixel rotational diagrams for the detected CH$_3$OH transitions \citep[see Figure \ref{fig:rotdiags}]{Goldsmith&Langer1999}. %as shown in Figure \ref{fig:rotdiags}).

To compute $T_{rot}$, we begin with the equation of \nupper,
\begin{equation}
    N_u=\frac{N_{tot}g_u}{Q_{rot}}\left[ \mathrm{exp}\left(\frac{E_u}{k_{B}T_{ex}}\right)\right]^{-1}
    \label{eq:gen_nupper}
\end{equation}
where $Q_{rot}$ is the rotational partition function \citep{MangumandShirley2015}.%$N_{tot}$ is the total molecular column density and 

Rearranging (\ref{eq:gen_nupper}) and setting $T_{ex}=T_{rot}$, we use
\begin{equation}
    \mathrm{log}_{10}\left(\frac{N_u}{g_u}\right)=\frac{-E_u}{k_{B}T_{rot}}\mathrm{log}_{10}(e)+\mathrm{log}_{10}\left(\frac{N_{tot}}{Q_{rot}}\right)
\end{equation}
 to calculate $T_{rot}$ and construct maps of the gas temperature within a region of interest.% as shown in Figure \ref{fig:texmaps}. 
 
 To compute the error on $T_{rot}$ and $N_{tot}$, we initially compute the statistical error on the linear fit as derived by \citet{Hogg+2010}. However, this method produces $T_{rot}$ error ($\sigma_{T_{rot}}$) values that do not appropriately capture the uncertainty implied by the observed scatter in the rotational diagrams (e.g., $\sigma_{T_{rot}} \lesssim 10$ K). {Some methanol transitions (e.g., the 5-4 line, see the second datapoint in Figure \ref{fig:rotdiags}) show $N_u/g$ values that consistently deviate from LTE predictions. We do not explore the apparent non-LTE excitation of these lines in this work. However, to account for the impact these deviations may have on our results,} we employ a simple bootstrapping method using the \texttt{astropy.stats.bootstrap} package. {In each pixel, we perform 1000 bootstrap resampling iterations. We refit the LTE model to each of these samples, and find that bootstrap samples are reasonably approximated by a Gaussian distribution. We therefore take the standard deviation of the resulting $N_{tot}$ and $T_{rot}$ values to obtain the uncertainty estimate. We report these bootstrap errors in Table \ref{tab:measuredprops} and propagate them through all values derived from $T_{rot}$ and $N_{tot}$.}

 \subsubsection{The effects of \methanol opacity on $T_{rot}$}\label{subsec:opacity}
{As the density of \methanol increases, the lower-J transitions will become optically thick and self-absorbed, causing the observed $N_{tot}$ from the rotational diagrams to stop growing and the observed $T_{rot}$ to increase. However, as N(H$_2$) may continue to independently increase, this can artificially decrease X(\methanol). We therefore set out to test the influence of line optical depth on our results.}

\begin{figure}
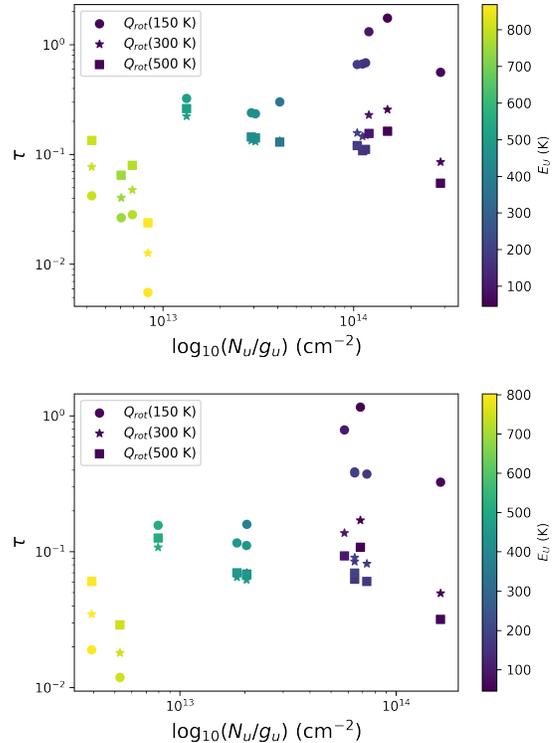

    \centering
    \includegraphics[scale=0.5]{f4.pdf}
    \includegraphics[scale=0.5]{f5.pdf}
    \caption{{Line opacities for fiducial hot core temperatures of 150 K (circles), 300 K (stars), and 500 K (squares) for \sgrbtwosouth (upper) and \dsii (lower). The color of the data points corresponds to the $E_u$ value for each transition.}}
    \label{fig:lineopacities}
\end{figure}

{Qualitatively examining the spectra from pixels showing this abundance decrease, we only find evidence of \methanol self-absorption in the 4-3 line in \dsi, suggesting that $\tau_{CH_3OH}$ is not extremely high across the sample. To quantitatively characterize the influence of $\tau_{CH_3OH}$ on $T_{rot}$, we examine the line opacities over the pixel showing the highest $T_b$ in the 4-3 line of each source. These regions are coincident with the continuum peaks for every source except \sgrbtwosouth, in which case we test the highest dust continuum peak and the $T_{rot}$ peak. We use the measured $N_{tot}$ value in the selected pixel and the \texttt{lte\_molecule} module of \texttt{pyspeckit} to model the optical depth of each line over a range of fiducial hot core temperatures, as shown in Figure \ref{fig:lineopacities}.}

{These models show a trend of increasing $\tau$ with \nupper, with the highest values of $\tau$ reaching approximately 0.3 for the highest-opacity sources. To produce marginally optically thick lines consistent with our \nupper measurements (i.e., $\tau \gtrsim 0.6$), the measured $T_{rot}$ in the chosen pixel would need to be overestimated by a factor of 2 or more. Assuming we were underestimating the opacity in the lowest-$J$ lines such that the true value of $\tau$ is of order 0.6, we apply the optical depth correction to our \nupper measurements as given by \citet{Goldsmith&Langer1999}. We find that these modeled values of \nupper differ from the values we measure with the optically thin approximation by 28.3\%, equivalent to no more than a 10 K increase in the $T_{rot}$ error we measure.}

{As an additional test, we are able to reasonably reproduce the spectra for the 23 identified species using our measured temperatures and estimates on their total column densities ({see Section \ref{sec:temps}}). Taken together, these results show that $\tau_{CH_3OH}$ does not significantly impact our measured temperatures and therefore is not a dominant factor in our analysis.}

\section{Results} \label{sec:results}

The general physical properties of each examined hot core are shown in Table \ref{table:obsprops}. Source radii were {measured as} the radius at which the azimuthally average temperature reached {its minimum value or} 150 K ($R_{{core}}$). {We choose 150 K to be consistent with predictions for the sublimation temperatures of \methanol and frequently studied COMs with water ice ( $\sim120-160$ K, see \cite{Garrod+2022}) and because} we do not observe temperatures below 100 K in several cores, unlike what is found in canonical Galactic disk hot cores (e.g., \cite{Gieser+2021}). Uncertainties on the radius were taken as the $1\sigma$ beam radius. {Source masses were measured by computing the sum of the mass within each target's \textit{R$_{core}$}.}

\begin{figure*}
    \centering
    \includegraphics[scale=0.5]{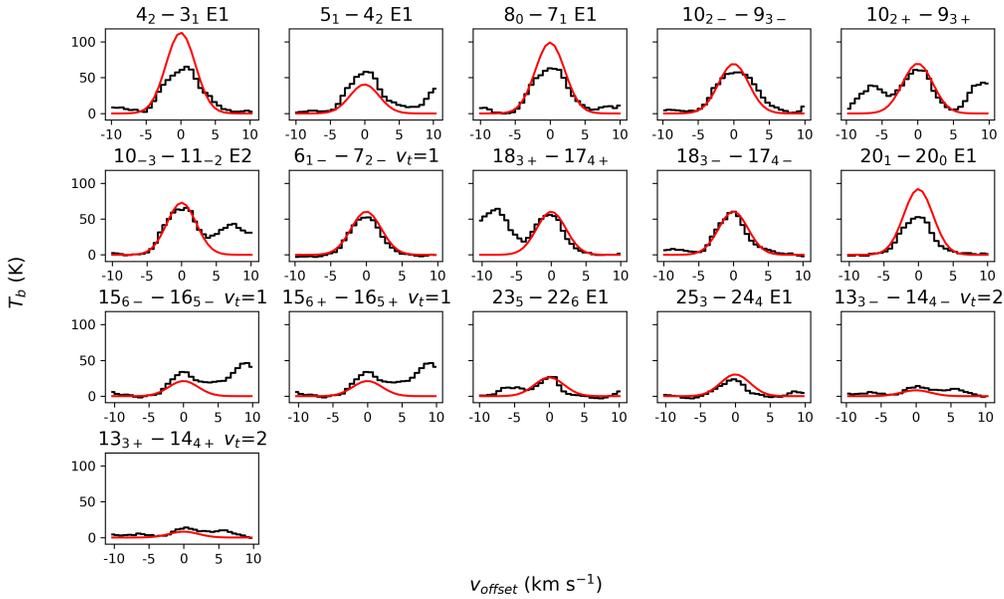}
    \caption{{Spectra of the \methanol lines used to produce the temperature maps for \sgrbtwosouth from its representative pixel. Black lines are the ALMA 1 mm measurements, and red lines are LTE model lines computed using our measured $T_{rot}$, $N_{tot}$, and line widths. The model reproduces the data well, with some deviation from predicted brightnesses and slight offsets from the systemic velocity of the source. These variations are present with different lines across the data set, which we attribute to limitations of our single-temperature, single-velocity assumption.}}
    \label{fig:multilineplot}
\end{figure*}

{Figure \ref{fig:multilineplot} shows the \methanol lines used to produce the temperature map for \sgrbtwosouth and overplotted \texttt{pyspeckit}-modeled lines produced using the measured $T_{rot}$, $N_{tot}$, and line width in this pixel. The model reproduces the data well, with deviation from the model line brightnesses in some lines and slight velocity offsets in others. In some other sources, the models less convincingly reproduce the data, overpredicting the line brightnesses by factors of 2 to a few (e.g., Figure 17.3, see Appendix \ref{app:multilineplots}). We attribute these deviations to limitations of our single-temperature, single-velocity assumption, as it is likely that the temperature and velocity structure of the cores changes on smaller scales.}

\begin{deluxetable*}{cccccc} \label{table:obsprops}

\tablecaption{Observational properties of the \deepsouth hot cores. Uncertainties on each core's RA and DEC are given in parentheses and equivalent to one beam width. Velocities are the line of sight velocity measured from each source's representative line.}

\tablehead{\colhead{Source Name} & \colhead{RA (ICRS (J2000))} & \colhead{DEC (ICRS (J2000))} & \colhead{Systemic Velocity} & \colhead{Representative Line}\\ 
\colhead{} & \colhead{(hh:mm:ss.ss)} & \colhead{(dd:mm:ss.ss)} & \colhead{(km s$^{-1}$)} & \colhead{}}

\startdata
\sgrbtwosouth & 17:47:20.48(0.02) & -28:23:46.3(0.3) & {69.3} & $20_{1}-20_{0}$ E1\\
\dsi & 17:47:19.58(0.02) & -28:23:49.9(0.3) & {56.2} & $8_{0}-7_{1}$ E1\\
\dsii & 17:47:20.05(0.02) & -28:23:46.7(0.3) & {48.7} & $8_{0}-7_{1}$ E1\\
\dsiii & 17:47:19.99(0.02) & -28:23:48.9(0.3) & {52.8} & $10_{2-}-9_{3-}$ \\
\dsiv & 17:47:19.77(0.02) & -28:23:43.5(0.3) & {54.6} & $20_{1}-20_{0}$ E1 \\
\dsv & 17:47:19.71(0.02) & -28:23:51.6(0.3) & {55.1} & $8_{0}-7_{1}$ E1\\
\dsvi & 17:47:21.12(0.02) & -28:24:18.3(0.3) & {49.8} & $8_{0}-7_{1}$ E1\\
\dsvii & 17:47:22.23(0.02) & -28:24:34.0(0.3) & {48.7} & $8_{0}-7_{1}$ E1\\
\dsviii & 17:47:22.04(0.02) & -28:24:42.6(0.3) & {49.8} & $8_{0}-7_{1}$ E1\\
\dsix & 17:47:23.46(0.02) & -28:25:52.1(0.3) & {47.3} & $8_{0}-7_{1}$ E1\\
\enddata

\end{deluxetable*}

In Figure \ref{fig:transitionmaps}, we show maps of the number of detected \methanol transitions ($n_{transition}$) per pixel for \sgrbtwosouth and \dsi. These maps are characteristic of the DS hot core population, showing excited \methanol that is both cospatial with regions of 1 mm dust continuum and spans several thousand AU away from the primary continuum features. Peak $n_{transition}$ values overlap with regions of highest continuum emission and are roughly constant across these same regions. $n_{transition}$ maps for the remaining hot cores are shown in Appendix \ref{app:ntransition}.

\begin{figure}
    \centering
    \includegraphics[scale=0.5]{f7.pdf}
    \includegraphics[scale=0.5]{f8.pdf}
    \caption{\methanol $n_{transition}$ maps for \sgrbtwosouth (upper) and \dsii (lower). Black contours show the 1mm dust continuum emission at 3, 6, 8, 12, and 32$\sigma$. {Cyan stars mark the position of the represenative pixel for each source, and the cyan circles mark $R_{core}$.} Both cores show \methanol emission spanning several thousands AU away from the continuum, with peak $n_{transition}$ values overlapping with the continuum. Peak $n_{transition}$ values are broadly constant across the continuum overlap regions. \sgrbtwosouth has a higher excitation state than \dsii, based on $n_{transition}$, and this excitation extends almost uniformly throughout the main body of the core. In addition, there are regions to the south which are also highly excited. \dsii shows a similar uniform excitation throughout its extent.}
    \label{fig:transitionmaps}
\end{figure}

\begin{deluxetable*}{ccccccc}\label{tab:measuredprops}

\tablecaption{Measured properties for the \deepsouth hot cores.}

\tablehead{\colhead{Core} & \colhead{$T_{peak}$} & \colhead{$N_{tot,max}^a$} & \colhead{$R_{core}$} & \colhead{$N_{H_2,max}^a$} & \colhead{$M_{core}$} & \colhead{X(\methanol)$_{peak}$} \\ 
\colhead{} & \colhead{(K)} & \colhead{($\times 10^{17}$ cm$^{-2}$)} & \colhead{(AU)} & \colhead{($\times10^{24}$ cm$^{-2}$)} & \colhead{(M$_{\odot}$)} &  \colhead{($\times10^{-7}$)} } 

\startdata
DS1 & $306\pm75$ & $14.22\pm0.45$ & $5004\pm1489$ & $1.70\pm0.03$ & $874\pm67$ & $8.44\pm1.61$ \\
DS2 & $239\pm41$ & $5.81\pm0.18$ & $3753\pm1489$ & $0.69\pm0.11$ & $423\pm31$ & $8.54\pm1.48$ \\
DS3 & $351\pm47$ & $6.49\pm0.62$ & $5838\pm1489$ & $1.83\pm0.32$ & $1568\pm104$ & $3.8\pm0.7$ \\
DS4 & $398\pm57$ & $10.68\pm1.06$ & $4881\pm1489$ & $1.35\pm0.21$ & $843\pm65$ & $7.74\pm1.42$ \\
DS5 & $357\pm90$ & $1.75\pm0.16$ & $3934\pm1489$ & $0.62\pm0.10$ & $420\pm36$ & $3.5\pm0.8$ \\
DS6 & $427\pm69$ & $12.33\pm0.21$ & $6672\pm1489$ & $2.22\pm0.47$ & $2192\pm125$ & $6.11\pm1.18$ \\
DS7 & $282\pm51$ & $4.7\pm0.18$ & $4107\pm1489$ & $1.37\pm0.30$ & $612\pm51$ & $3.4\pm0.8$ \\
DS8 & $300\pm42$ & $3.59\pm0.41$ & $3587\pm1489$ & $0.69\pm0.10$ & $287\pm25$ & $5.31\pm1.0$ \\
DS9 & $252\pm39$ & $2.42\pm0.14$ & $3956\pm1489$ & $0.38\pm0.06$ & $203\pm16$ & $6.49\pm1.12$ \\
SgrB2S & $662\pm191$ & $23.56\pm0.62$ & $9436\pm1489$ & $3.01\pm0.60$ & $4842\pm787$ & $11.31\pm1.76$ \\
\enddata

\tablenotetext{$a$}{Peak $N_{tot}$ values and peak $N_{H_2}$ are pulled from different pixels.}

\end{deluxetable*}

\subsection{Temperatures} \label{sec:temps}
Examples of rotational diagrams from \sgrbtwosouth and \dsi are shown in Figure \ref{fig:rotdiags}. Peak temperatures for the full sample are shown in Table \ref{tab:measuredprops}. Sample temperature maps from \sgrbtwosouth and \dsii are shown in Figure \ref{fig:texmaps}, with those of the remaining sources shown in Appendix \ref{app:texmaps}. {All temperature maps only include temperatures $T_{rot}\geq3\sigma_{T_{rot}}$.} A majority of the cores (seven of 10) show temperature peaks within one beam width of the core's continuum peak. In \sgrbtwosouth, local temperature peaks are similarly present over its constituent continuum peaks. In \dsvii and \dsix, hot ring features surround the continuum peaks.

{We note that the 5-4 transition is higher than our LTE model predicts (see e.g., the second data point in both panels of Figure \ref{fig:rotdiags}). %We do not explore the excitation of this line in this work, but we are able to rule out line blending as the cause of this behavior. The nearest identified contaminants (the $23_{2,22}-22_{2,21}$ F=22-22 line of CH$_2$CHCN and the $20_{1,20}-19_{0,19}$ line of CH$_3$OCHO) are $-12$ and 29 km s$^{-1}$ away and do not overlap with the 5-4 line.
The effects of this line are accounted for in the bootstrapping uncertainty estimation we employ (see Section \ref{subsec:trotmeasurement}), as removing it produces less than a 1$\sigma$ difference in our temperatures. Similar plots for the remaining hot cores are shown in Appendix \ref{app:multilineplots}.}

{Appendix \ref{app:spectra} shows continuum-subtracted ALMA spectra of all four spectral windows from the representative pixel of the observed hot cores.} Overplotted are \texttt{pyspeckit}-modeled lines produced by \methanol and a sample of identified species at the measured $T_{rot}$(\methanol) in the chosen pixels. We select these sources to illustrate the range of line densities and brightnesses present within the sample. {We assume all molecules have a common $T_{rot}$ based on the qualitatively similar line widths in various species, which we interpret to mean they are coming from similar parcels of gas in the envelope. At the same time, we note that this is meant as an initial estimate.} Based on these assumptions, we identify 25 unique molecular species (including isotopologues), and one vibrationally excited species (HC$_3$N v$7=1$). Among the identified species, 11 are COMs: \methanol, $^{13}$\methanol, $^{13}$CH$_3$CN, CH$_3$CHO, CH$_3$NCO, CH$_3$OCHO, CH$_3$OCH$_3$, C$_2$H$_5$OH, NH$_2$CHO, CH$_2$CHCN, and CH$_3$CH$_2$CN. While many lines remain unidentified, the hot cores show a similar distribution of species across the identified COMs, as eight of these 10 species are detected in each core. Assumed total column densities for each species are manually tuned to match the data, and there is good agreement ($T_b$ within factors of a few) between the models and the dataset.

With these detections, we now have reasonable templates for hot cores that we can use for further, deeper searches in the data.

\begin{figure}
    \centering
    \includegraphics[scale=0.5]{f9.pdf}
    \includegraphics[scale=0.5]{f10.pdf}
    \caption{\methanol rotational temperature maps for \sgrbtwosouth (upper) and \dsii (lower). White contours show the 1 mm dust continuum and the contour levels are 3$\sigma$, 6$\sigma$, 8$\sigma$, 16$\sigma$, and 32$\sigma$ ($\sigma=0.2$ mJy beam$^{-1}$). The hot gas within \sgrbtwosouth spans a large area ($\sim$0.5 pc across its horizontal) and is dominated by the distinct hotspot in its northeast. \dsii shows a more uniform morphology, with an extended envelope of warm ($\sim200$ K) gas surrounding its continuum. Temperature maps for the remaining cores can be found in Appendix \ref{app:texmaps}.}
    \label{fig:texmaps}
\end{figure}

\subsection{Core physical structure}
{As noted in Section \ref{sec:intro}, Sgr B2 provides a valuable laboratory within which to explore the resolved properties of stars forming in conditions resembling those found near the peak of cosmic star formation. In detail, the structural properties (i.e., the thermal and density gradients) of CMZ's massive protostars are relatively unexplored. These properties are of particular importance, as they inform modeling of the physical and chemical evolution of massive protostars (see e.g., \citep{vanderTak+2000}). To begin investigating these quantities in our sample, we follow \citet{Gieser+2021}, who created radial power-law indices to model the physical characteristics of a catalogue of hot cores in the Galactic Disk. In the following subsection, we briefly discuss radial temperature and density profiles and power-law fits we created to facilitate comparison between Deep South and this catalogue (see Section \ref{subsec:diskcomparison}).}
\subsubsection{Radial temperature profiles}
\begin{figure}
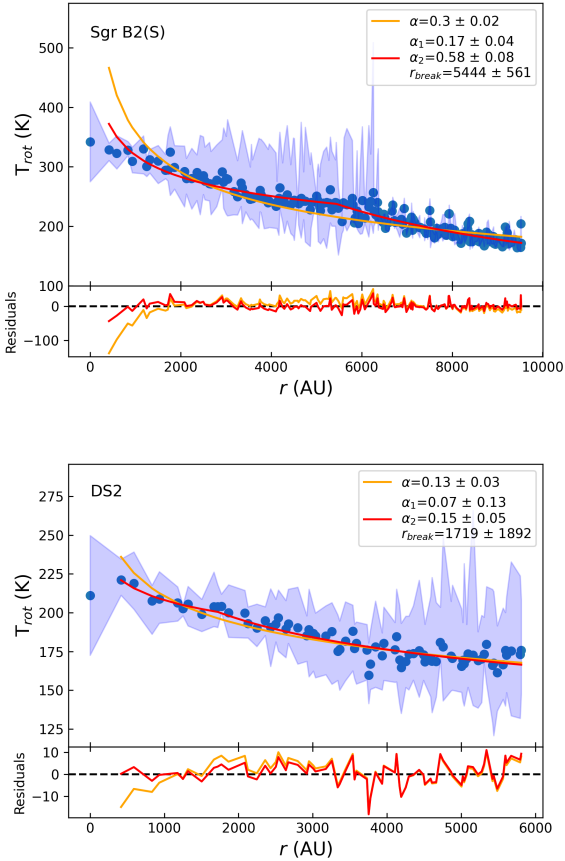

    \centering
    \includegraphics[scale=0.5]{f11.pdf}
    \includegraphics[scale=0.5]{f12.pdf}
    \caption{Radial temperature profiles for \sgrbtwosouth (upper) and \dsii (lower). As \sgrbtwosouth contains multiple continuum sources, its profile is centered on its temperature peak. \dsii is centered on the peak of its continuum. Both cores show clear peaks indicative of central heating, however \sgrbtwosouth has a significantly warmer temperature peak. Overplotted are single (orange) and two-component broken (red) power law fits as described in Section \ref{sec:temps}. The residuals to the power law fits are shown in the lower panels in each plot.}
    \label{fig:radtempprof}
\end{figure}

 We constructed radial temperature distributions for each core, with those of \sgrbtwosouth and \dsi shown in Figure \ref{fig:radtempprof}. We initially fit single power-law profiles to these distributions, as described by the following equation:

\begin{equation} \label{eq:singlepowerlaw}
    T_{rot}(r)=A\left(\frac{r}{r_{break}}\right)^{-\alpha}
\end{equation}

where $A$ is the normalization constant, $r$ is the radial distance from the hot core continuum peak, $r_{break}$ is the break point of the power law, and $\alpha$ is the single power-law index. However, these profiles are a poor fit to the data for $r \lesssim 4000$ AU. To better parameterize the data, we fit broken power-law profiles to the data using the following equation: 
\begin{equation}
    T_{rot}(r)=
        \begin{array}{cc} % 11 was giving an error here
            A\left(\frac{r}{r_{break}}\right)^{-\alpha_1} & \quad :r < r_{break} \\
            A\left(\frac{r}{r_{break}}\right)^{-\alpha_2} & \quad :r > r_{break}
        \end{array}
\end{equation}
where $\alpha_1$ and $\alpha_2$ are the interior and exterior power-law indices, respectively. We find that broken power laws better describe the behavior towards the interiors of the hot cores, where the temperatures initially fall slower than single power laws predict. %Exceptions are potentially DS4 and DS9, which misbehaved. Will have to revisit those cores to get the fitter to cooperate

Table \ref{table:fittedparams} reports the parameters of the broken power-law fits for each hot core in the sample. The temperature distributions for hot cores \dsii through \dsix are shown in Appendix \ref{appendix:texprofs}. We compare $\alpha_2$ to similar measurements performed on in the Galactic Disk CORE survey in Section \ref{subsec:diskcomparison}.

\begin{deluxetable*}{ccccc} \label{table:fittedparams}

\tablecaption{Fitted parameters of the radial temperature profiles for each hot core. %$\alpha_1$ and $\alpha_2$ are the power law indices measured in the first and second components of the profiles.
}

\tablehead{\colhead{Core} & \colhead{$\alpha_1$} & \colhead{$\alpha_2^\dagger$} & \colhead{$r_{break}$} & \colhead{$p$}} 

\startdata
DS1 & $0.15\pm0.03$ & $0.36\pm0.01$ & $1931\pm224$ & $1.9\pm0.008$ \\
DS2 & $0.07\pm0.04$ & $0.15\pm0.01$ & $1719\pm547$ & $1.06\pm0.01$ \\
DS3 & $0.04\pm0.08$ & $0.31\pm0.03$ & $1533\pm362$ & $1.64\pm0.01$ \\
DS4 & $0.11\pm0.06$ & - & - & $1.886\pm0.008$ \\
DS5 & $0.18\pm0.17$ & $1.17\pm0.11$ & $2637\pm331$ & $1.26\pm0.02$ \\
DS6 & $0.1\pm0.2$ & $0.36\pm0.02$ & $1029\pm439$ & $1.689\pm0.003$ \\
DS7 & $-0.05\pm0.07$ & $0.65\pm0.03$ & $1844\pm113$ & $1.714\pm0.008$ \\
DS8 & $0.1\pm0.09$ & $0.55\pm0.08$ & $1943\pm323$ & $1.72\pm0.01$ \\
DS9 & $0.55\pm0.08$ & $0.31\pm0.08$ & $4293\pm895$ & $1.65\pm0.01$ \\
Sgr B2(S) & $0.17\pm0.04$ & $0.58\pm0.08$ & $5444\pm561$ & $2.05\pm0.01$ \\
\enddata

\tablenotetext{\dagger}{We set $\alpha_2=q$ in the lower panel of Figure \ref{fig:pindex}, except in \dsiv, {\dsix, and \sgrbtwosouth.} We use $\alpha_1$ for {these} sources as the temperature profile is well described by this value until relatively large radii (i.e. $r_{break}$ for this source). {In \dsiv, its measurements of $\alpha_2$ and $r_{break}$ are not statistically significant, so we omit them with dashes.}}

\end{deluxetable*}

\subsubsection{Radial density profiles}\label{subsec:radialdensityprofiles}
{To compute densities for our sample, we first set the dust temperature  $T_{dust}=T_{rot}$. While the dust temperatures are not well known at these scales, they should be thermally coupled to the gas (\cite{Clark+2013}) in the expected density regime  of the hot cores ($n\gtrsim10^7$ cm$^{-3}$; e.g., \cite{Motte+2018}). We then compute the peak dust optical depth for each source using the following equation:}
\begin{equation}
    \tau=\frac{S_{\nu}c^2}{2k_BT_{rot}\nu^2}
\end{equation}
{where $S_\nu$ is the dust surface brightness in Kelvin, $c$ is the speed of light, and $\nu$ is the representative frequency of the continuum. The highest value of $\tau$ among the hot cores is 0.07, meaning optically thin dust. We then derive H$_2$ column densities $N(\mathrm{H}_2)$. Using the continuum brightness temperatures $T_B$ and again setting$T_{dust}=T_{rot}$, we arrive at the following equation:}
\begin{equation}
    N(\mathrm{H}_2)=\frac{T_b \Omega D^2 c^2}{2 k_B \nu^2 \kappa_\nu \mu A_{beam} T_{rot}}
\end{equation}
{where $\Omega=2.79\times10^{-12}$ sr is the beam solid angle, $D=8.34$ kpc is the distance to Sgr B2 , $c$ is the speed of light, $\nu=217.35$ GHz is the representative frequency of the continuum, mean molecular weight $\mu=2.8$ \citep{Kauffman+2008}, and $A_{beam}$ is the beam area in AU at the distance of Sgr B2. We choose a distance of 8.34 kpc to Sgr B2 to maintain consistency with \citet{Schmiedeke2016} (see Section \ref{subsec:youngermassive}). The opacity constant $\kappa_\nu=0.008$ cm$^2$ g$^{-1}$ \citep{OssenkopfandHenning1994} implicitly assumes a gas-to-dust ratio of 100. We include N($H_2$) maps for each source in Appendix \ref{app:nh2maps}.}

\begin{figure}
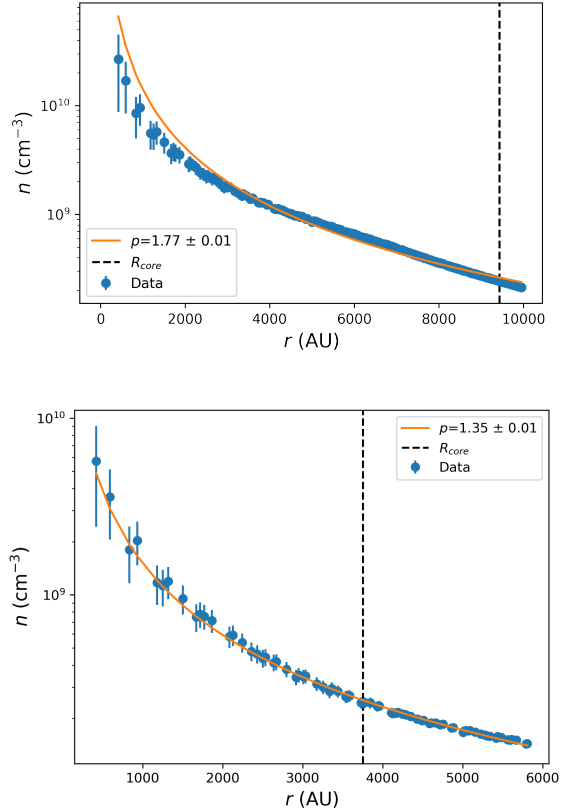

    \centering
    \includegraphics[scale=0.52]{f13.pdf}
    \includegraphics[scale=0.5]{f14.pdf}
    \caption{Radial density profiles for {\sgrbtwosouth} and \dsii. The orange line is our single power law fit for each source, and we report the associated power law index $p$ using the same nomenclature as \citet{Gieser+2021}. The vertical black line shows $R_{core}$ for each source.}
    \label{fig:radialdensityprofile}
\end{figure}

Using our N(H$_2$) measurements, we create radial density profiles assuming spherical symmetry for each hot core {(e.g., Figure \ref{fig:radialdensityprofile})}. We fit single power law profiles to each distribution following Equation \ref{eq:singlepowerlaw} and report the fitted power law index $p$ in Table \ref{table:fittedparams}. Peak densities in all cores are approximately 10$^{10}$ cm$^3$. {We note here that lower dust temperatures would imply a higher optical depth. However, this would act to increase the densities we measure or make them lower limits, if the cores were optically thick.} This implies that $T_{dust}$ is strongly coupled to the gas temperatures {in our sources} (e.g., \cite{Clark+2013}). In these conditions and at our resolution, the dust within the hot cores is optically thin with values $\tau<0.07$. We discuss $p$ and how it compares between DS and a selection of Galactic Disk hot cores in Section \ref{subsec:diskcomparison}.

\subsection{Abundances}\label{subsec:abundresult}
\begin{figure}
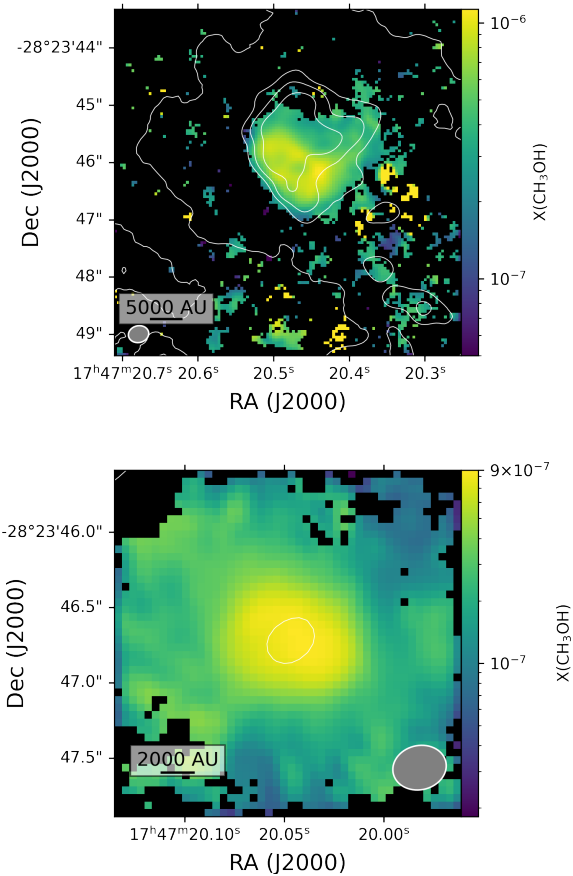

    \centering
    \includegraphics[scale=0.6]{f15.pdf}
    \includegraphics[scale=0.6]{f16.pdf}
    \caption{\methanol abundance maps for \sgrbtwosouth (upper) and \dsii (lower). Regions in which the 1 mm continuum (white contours) is detected at less than $3\sigma$ have been masked out.} %Additionally, regions of high molecular complexity (i.e. rotational diagrams with large scatter) are also masked per the bootstrapping method described in \ref{sec:methods}. Both hot cores show lower abundances coincident with local peak in the continuum. (To do: fix contours in the upper plot, they're out of date)}
    \label{fig:abundancemaps}
\end{figure}

{Using the N($H_2$) measurements described in Section \ref{subsec:radialdensityprofiles}, we create methanol abundance (X(\methanol)) maps} by taking the ratio $N_{tot}$/$N(H_2)$ {(see Table \ref{table:obsprops} for peak $N_{tot}$ and $N(H_2)$ values)}. Abundance maps for \sgrbtwosouth and \dsii are shown in Figure \ref{fig:abundancemaps}, and those of the other sources are shown in Appendix \ref{app:abundmaps}.

The hot core \sgrbtwosouth contains an HII region in its northern regions (e.g., \cite{Lu+2019,Meng+2022}). As free-free contamination of the continuum is a concern in this region, we omit the HII region when measuring the abundances in this source. There are no free-free detections in the remaining hot cores.

\begin{figure*}
    \centering
    \includegraphics[scale=0.63]{f17.pdf}
    \includegraphics[scale=0.63]{f18.pdf}
    \caption{Abundance vs. temperature plots for \dsii (left) and \sgrbtwosouth (lower) showing the variation in abundance behavior across sources. {Each point represents a single pixel.} In \dsi, the abundance increases monotonically with temperature until the temperature peak is reached. In \sgrbtwosouth, X(\methanol) peaks near 210 K before descending to a constant value at higher temperature. The abundance vs temperature plots for the remaining hot cores can be found in Appendix \ref{app:abundvstex}.}
    \label{fig:abundancevstemperature}
\end{figure*}

{We qualitatively identify a trend of increasing abundance with $T_{rot}$ across the sample. However, there is variation on a source-by-source basis: we do not observe a common peak abundance value, or a common temperature at the abundance peak in any hot core. Temperatures at the abundance peaks range from $\sim200$ to 400 K. A majority (\dsi-\dsix) of the sources show monotonic increases in abundance until the highest measured temperatures. However \sgrbtwosouth (upper panel in Figure \ref{fig:abundancevstemperature}) is unique in the sample, as its abundance rises to a peak at 210 K, and then flattens to a roughly constant value out to the highest temperatures for which we have abundance measurements. The individual X(\methanol) vs $T_{rot}$ distributions of the remaining hot cores are shown in Appendix \ref{app:abundvstex}.}

\begin{figure}
    \centering  \includegraphics[scale=0.45]{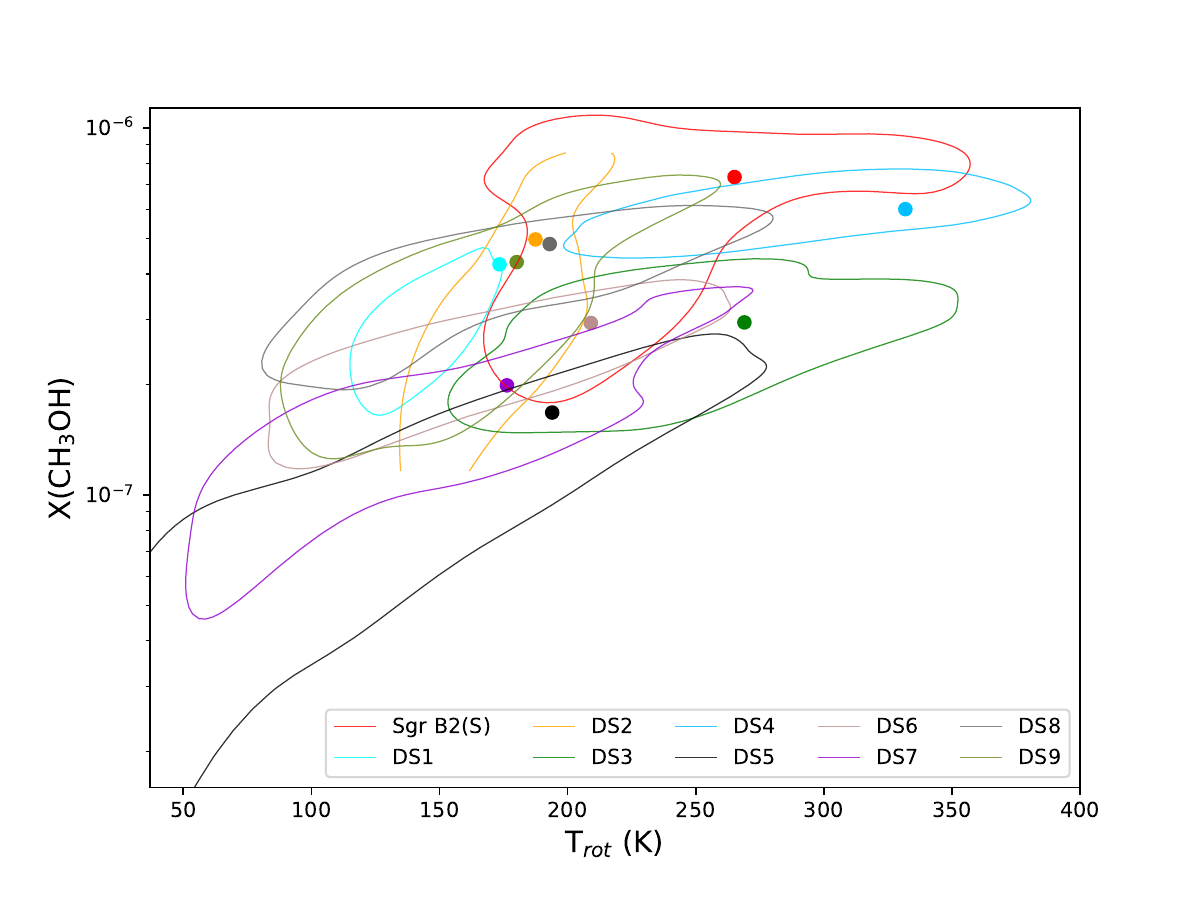}
    \caption{Diagnostic plot showing the distribution of X(\methanol) as a function of $T_{rot}$ for the DS hot cores. {Contours show the distribution of abundances in each hot core as a function of temperature. They show a positive correlation with temperature, with scatter attributed to core-specific chemical evolution.} Colored points correspond to the signal-to-noise-weighted averages, {where the weights are the signal-to-noise of the abundance} for each hot core.} %Contours s X(\methanol) shows roughly an order of magnitude scatter in possible values at low temperatures (100-300 K), indicating a relatively poor correlation with $T_{rot}$.} 
    \label{fig:abundancesummary}
\end{figure}

To compare the abundance behavior of each DS hot core in the sample, we plot a kernel density estimate of the abundances and temperatures for each source in Figure \ref{fig:abundancesummary}. The increasing abundance trend is visible here, with scatter in the distribution arising from the unique abundance behavior in each hot core. {We briefly discuss potential physical drivers for these results and compare the peak abundances in Deep South to those of hot cores in other regions of Sgr B2 and the Galaxy in Section \ref{subsec:xch3ohingalacticcontext}}.

\begin{figure}
    \centering
    \includegraphics[scale=0.5]{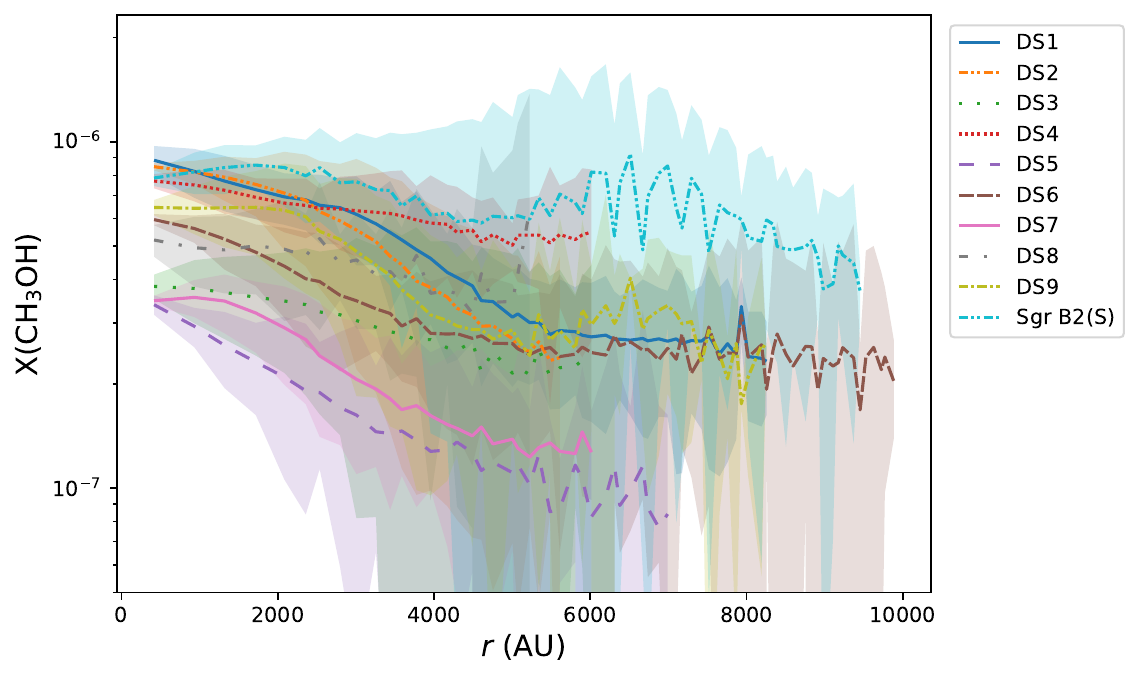}
    \caption{Radial abundance profiles for the full sample of hot cores in this work.  The profiles were smoothed with a median kernel of 3. Shaded regions show the $1\sigma$ distribution of abundances for each radial bin. {All sources have a decreasing abundance distribution out to large radii, consistent with typical models of hot cores.}}
    \label{fig:radialabundancedrop}
\end{figure}

In Figure \ref{fig:radialabundancedrop}, we show radial abundance profiles for the full hot core dataset.  {In all of our sources, we identify a trend of decreasing \methanol abundance with separation from the dust continuum cores, with some scatter depending on the abundance morphology of individual sources. This is broadly consistent with the canonical image of hot cores, where thermal feedback from embedded protostars sublimates the grain mantles in the protostellar envelope.} %The source  We discuss biases that may affect these results and potential physical mechanisms in Section \ref{subsec:xch3ohingalacticcontext}.

\section{Discussion} \label{sec:discussion}
{The hot cores of \deepsouth are a heterogeneous population with a range of temperatures, masses, abundances, and structural properties. In the following sections, we discuss these sources in the context of other hot core populations in the Galaxy, including those within Sgr B2, and what our results suggest about the recent epoch of star formation in Sgr B2.}

\subsection{\sgrbtwosouth: a younger Sgr B2(N) and (M)?}\label{subsec:youngermassive}

\citet{Schmiedeke2016} use three-dimensional modeling of archival millimeter and FIR datasets from the Very Large Aarray, APEX, Submillimeter Array, and Herschel to derive a stellar cluster radius of 0.35 pc for \sgrbtwosouth. Within this radius, we quote their measured H$_2$ mass of 4472 M$_\odot$. Our higher-resolution observations in the millimeter regime reveal a cluster of hot cores neighboring \sgrbtwosouth within this same radius (sources \dsi through \dsv), which we have named the Deep South Hot Core Complex. The newly discovered hot cores contain a total molecular gas mass of {4128} M$_\odot$, yielding combined molecular gas mass with \sgrbtwosouth ($M_{core}={4842}$ M$_{\odot}$) of {8970} M$_\odot$. These measurements show that \sgrbtwosouth dominates the flux at 3 mm and 1 mm flux in this region. However, it {only} contains {just over half (54\%)} of the total high-mass star-forming mass. The remaining {46\%} is located in the cooler, less luminous complex members (\dsi-\dsv).

These results {nearly} double the mass in high-mass stars in the vicinity of \sgrbtwosouth, and there are numerous currently uncatalogued lower-mass or less evolved cores distributed in the same region. We therefore suggest that, instead of \sgrbtwosouth being a less massive sibling to Sgr B2(N) and (M), it and the neighboring hot cores throughout the complex may instead comprise a younger, comparably massive sibling.

\subsection{Comparison to Galactic disk hot cores}\label{subsec:diskcomparison}
We compare the properties of the DS hot core population to a selection of Galactic disk hot cores studied at comparable ($\sim3000$ AU) resolution in WB 89789, \citep{Shimonishi+2021} and in the CORE survey \citep{Gieser+2021}, and at comparable temperatures in W51 \citep{Goddi+2020}. The masses reported in Table 3 of \citet{Gieser+2021} were derived using a gas to dust ratio $\frac{M_{gas}}{M_{dust}}=150$ and an opacity constant $\kappa=0.009$ cm$^2$ g$^{-1}$ at 1.3 mm. These differ from the values we assume in this work ($\frac{M_{gas}}{M_{dust}}=100$, $\kappa=0.00858941$ cm$^2$ g$^{-1}$ at 1.3 mm; \cite{OssenkopfandHenning1994}), thus we scale their masses to allow for direct comparison to those of Deep South. The scaled masses differ from the values reported in \citet{Gieser+2021} by 8.9\%. %We additionally compute the inner number density $n_{in}$ and integrate the density profile of their cores out to the reported outer radii $R_{out}$ to arrive at the total mass contained within each hot core.

\begin{figure}
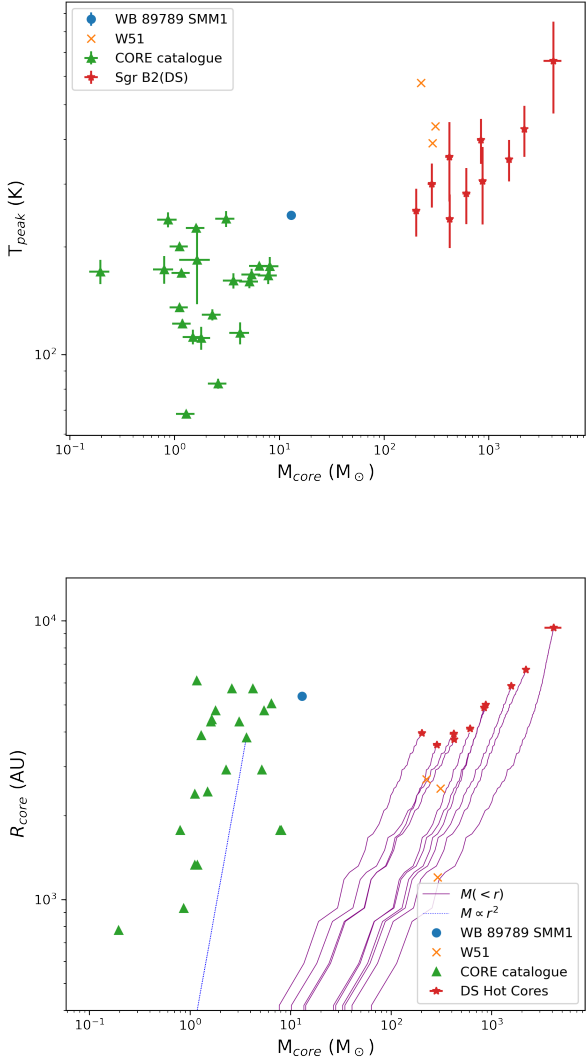

    \centering
    \includegraphics[scale=0.5]{f21.pdf}
    \includegraphics[scale=0.5]{f22.pdf}
    \caption{Comparison plots showing the distribution of peak temperatures (upper) and radii (lower) as a function of mass for the source studied in this work and a selection of Galactic disk hot cores which were studied at comparable resolution. The blue circle represents hot core WB 89789 SMM1 \citep{Shimonishi+2021}, green triangles are hot cores in the CORE catalogue \citep{Gieser+2021}, yellow crosses are from the W51 region \citep{Goddi+2020}, and red stars are the DS hot cores. Purple lines show the mass interior to each radial bin for each DS hot core. {The blue dotted line is an $M\propto r^{2}$ profile as a reference.} }
    \label{fig:masscomparisons}
\end{figure}

In the upper panel of Figure \ref{fig:masscomparisons}, we plot the distribution of peak temperatures as a function of core mass. The DS hot cores have masses and temperatures that are similar to the W51 cores, however they are more massive and generally warmer than the rest of the comparison set. In the lower panel of Figure \ref{fig:masscomparisons}, we also find that the DS cores have radii that are broadly consistent with those of the Galactic disk. %\sgrbtwosouth, the datapoint in the upper right, is an exception. However, this is expected given that this source is more comparable to a hot core cluster rather than an individual hot core like the others. 

\begin{figure}
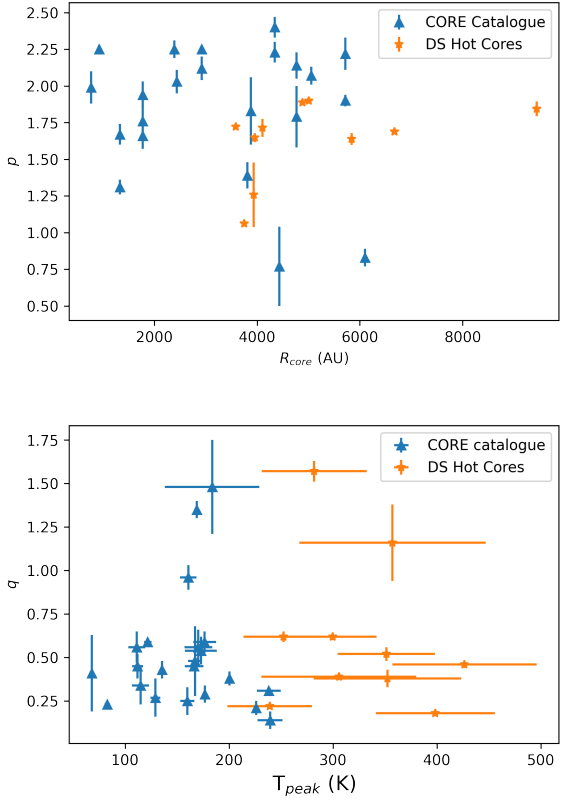

    \centering
    \includegraphics[scale=0.55]{f23.pdf}
    \includegraphics[scale=0.55]{f24.pdf}
    \caption{Plots comparing the fitted physical parameters of the CORE catalogue \citep{Gieser+2021} and the DS hot cores. Upper panel: distribution of density power law indices ($p$) as a function of $R_{core}$. Our sample shows an average shallower distribution of $p$ than their CORE counterparts. Lower panel: distribution of radial temperature indices ($q$) as a function of peak temperature. We compare $q$ to $\alpha_2$ in this work due to the resolution limits of the CORE survey. The DS hot cores have radial temperature index distribution that is {statistically indistinguishable from} the CORE catalogue.}
    \label{fig:pindex}
\end{figure}

We compare the measured radial density profiles $p$ (Section \ref{subsec:radialdensityprofiles}) to similar measurements done on the CORE catalogue. We find that the DS hot cores have, on average, shallower ($p={1.66} \pm 0.01$) density profiles than the CORE catalogue ($p=1.85 \pm 0.09$, \cite{Gieser+2021}; see Figure \ref{fig:pindex}, upper panel). {This means that, at a given radius, the DS hot cores are typically denser than the CORE hot cores, which aligns with the work of other authors concerning the threshold of star formation in the CMZ (see \citet{Henshaw+2023} for recent review). The CMZ shows evidence of a higher surface density threshold for star formation than canonical dense-gas relations predict (e.g., \cite{Gao&Solomon2004,Lada+2012}). The origins of this result are heavily debated due to its impact on the stellar initial mass function, but a clear answer has yet to be revealed. Qualitatively, however, it follows that protostars in the CMZ would show evidence of being denser than the disk if the surface density threshold for star formation is higher, as our observations demonstrate. For now, it is not clear if DS is representative of Sgr B2, the CMZ, or an outlier itself.}

{Performing the same comparison on the radial temperature power-law indices, we find that Deep South's hot cores are statistically indistinguishable from the CORE catalogue, with $p=0.44\pm0.04$ and $0.51\pm0.08$, respectively.} As \citet{Gieser+2021} did not observe a broken power-law behavior and our measured $\alpha_1$ only applies to the inner $\lesssim 3000$ AU of our cores, we compare $q$ to our $\alpha_2$ in the lower panel of Figure \ref{fig:pindex}. We use $\alpha_1$ in \dsiv, \dsix, and \sgrbtwosouth since their radial temperature profiles are well fit by this index out to large radii. {We can use this result to make inferences about the structure of the cores on scales smaller than our resolution. \cite{vanderTak+2000} use submillimeter continuum and molecular line observations and modeling to study the structural properties of a collection of massive YSOs in the Disk. Among the results in their work, they find radial temperature power-law indices which apply on comparable size scales to our work ($\gtrsim2000-3000$ AU) and with which our observations also agree. The overlap in these results lets us infer that the DS hot cores have increasingly optically thick interiors starting on the scale of $\sim1000$ AU and the thermal profile likely rises steeply on these size scales. Studying the temperatures on these size scales would require additional observations in Deep South, but the suggestion that the cores begin becoming optically thick on 1000 AU scales aligns with work published by \citet{Budaiev+2024} in Sgr B2(N). Higher-resolution, lower-frequency observations would likely allow us to begin probing this regime directly and elucidating the structure of these objects on the scales of their protostellar disks (e.g., \cite{BeltranDeWit2016}).}

{Taken together, in spite of the extreme environment of Sgr B2, we find that the hot cores of Deep South overall share structural properties with our chosen hot core comparison set. The sources are have similar radii and show evidence of having similar interior structures on the scales of their protostellar disks. However, they are also typically warmer than the comparison set, and their radial density power-law indices hint at the influence of the higher star formation threshold present in the CMZ. Future studies on YSOs in Deep South, Sgr B2(N), and elsewhere in the CMZ will help clarify if these results are representative of CMZ star formation or if Deep South is unique.}

\subsection{The relative age of Sgr B2(DS)}
The onset of star formation in Sgr B2 has been a point of interest in characterizing the star formation history of the CMZ and its broader geometry \citep[e.g.,][]{Kruijssen+2015,Henshaw+2016}, in addition to the physical and chemical history of Sgr B2 itself \citep[e.g.][]{Ginsburg2018,Bonfand2019, Meng+2022}. \citet{Meng+2022} place constraints on the relative timescale for star formation in Sgr B2 by observing the distribution of UCHII regions relative to dust sources throughout the cloud. They report that S and DS are likely the youngest regions in the cloud, as they contain the lowest number of UCHII regions (three) relative to N (eight) and M (40). By examining the characteristics of hot cores across the cloud, we explore an additional lens for determining relative age across Sgr B2, since hot cores represent a younger phase in massive protostellar evolution. {We find that \deepsouth is not at an obviously different evolutionary stage than N, with several cores sharing features with those of N.}

\citet{SanchezMonge+2017} identify 14 and 12 hot core candidates between Sgr B2(N) and (M), respectively (chemically ``rich" and ``rich?" sources in their Table 4). In N, three of their sources had been previously identified as hot cores: N1 (AN01) and N2 (AN02 and AN03, \cite{Belloche+2008,Belloche+2016}). Two more were confirmed and a new hot core was detected in observations by \citet{Bonfand2017}: N3 (AN08), N4 (AN14), and N5. The candidates in Sgr B2(M) have not yet been examined to the level of detail as the cores in N and in this work. %However, considering its more-evolved global characteristics \cite[e.g.,]{Qin+2011, Schmiedeke2016,Pols+2018}, it is likely that any confirmed hot cores in M are more evolved relative to those in N, S, and DS. 

\citet{Bonfand2017} proposed a tentative evolutionary sequence in which the presence of Class II \methanol masers (see \cite{Sobolev+1997}) and absence of outflows and UCHII regions may indicate less evolved sources. Investigating the outflows in DS is beyond the scope of this current work, but we are able to examine the presence of UCHII regions and masers. \sgrbtwosouth and N1 are likely the most evolved sources between these two regions, as they contain UCHII regions with no \methanol masers. Three DS hot cores contain masers and no UCHII region, as in N3 and N5, suggesting they may be at intermediate evolutionary stage: \dsii, \dsvi, and \dsviii \citep[e.g.,][]{HoughtonWhiteoak1995, Caswell1996, Lu+2019}. The remaining sources have neither UCHII regions nor \methanol masers, making their relative ages unclear using this sequence.

However, in addition to studying the outflows, we may learn more about their relative evolutionary state by examining masers in other species. \dsi and \dsii are associated with rare 4.83 GHz \formaldehyde masers (sources D and G in \citet{Mehringer1994}, reidentified as sources F4 and F5 in \citet{Lu+2019}). The pumping mechanism of this maser is still poorly understood, however recent results suggest it may be associated with high-mass star formation \citep{vanderWalt+2022}. As \dsii is a source that contains both a Class II \methanol maser and \formaldehyde maser, it presents a unique opportunity to study maser mechanisms, their relationship to outflows, and to use them to infer the evolutionary state of high-mass protostars.

\begin{figure}
    \centering
    \includegraphics[scale=0.5]{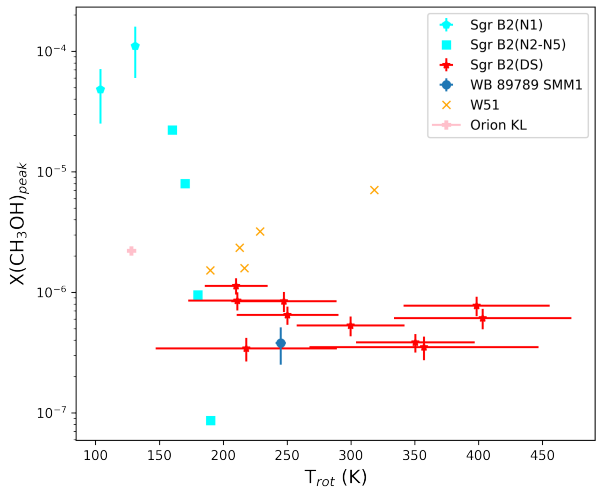}
    \caption{{Peak \methanol abundance (X(\methanol)$_{peak}$) versus conincident rotational temperature for the hot cores studied in this work and a selection of hot core populations in the CMZ and Galactic Disk. X(\methanol)$_{peak}$ occurs at a wide range of $T_{rot}$, with further variation within individual populations. There may be hints that peak abundance decreases with temperature.}}
    \label{fig:globalhotcore_xch3ohvstrot}
\end{figure}

{\subsection{Deep South's methanol abundances in the Galactic context}}\label{subsec:xch3ohingalacticcontext}
{As discussed in Section \ref{subsec:abundresult}, while we observe abundance increasing with temperature in every source, there are source-specific differences in peak abundance and peak temperature. To examine how these differences compare in the broader picture of Milky Way hot cores, we compare peak abundance values as a function of coincident temperature to a selection of hot cores in Sgr B2 (Sgr B2(N1), \cite{Busch+2022}; Sgr B2(N2-N5), \cite{Bonfand2017}), the Disk (W51, \cite{Ginsburg2017}; Orion KL, \cite{Crockett+2014}), and the outer Galaxy (WB 89789 SMM1, \cite{Shimonishi+2021}) in Figure \ref{fig:globalhotcore_xch3ohvstrot}. These cores were selected on the basis of having methanol abundance measurements at comparable spatial resolution ($\sim5000$ AU), which probes similar physical and chemical conditions to our work. We note that the two points from Sgr B2(N1) are taken from regions $\sim1$'' offset from the continuum peak in that source, as the peak of the continuum is optically thick at the measured frequencies ($\sim99$ GHz). \citet{Busch+2022} also employ two tracers to measure $H_2$ column density used in their abundances: dust continuum and C$^{18}$O. For consistency with our work, we use the abundances they measure using the dust-derived $H_2$ column density (see Table 3 in their work).} 

{The resultant distribution shows hints of a negative correlation between peak \methanol abundance and coincident $T_{rot}$, qualitatively consistent with predictions made in the co-desorption with water-ice model discussed in \citet{Garrod+2022}. %The measurements taken in Sgr B2(N1) show the highest peak abundances and are located at relatively low temperatures. The Orion KL hot core stands in contrast to Sgr B2(N1), with has an abundance that is roughly an order of magnitude lower at similar temperatures. Sgr B2(N2) and N3 occupy a slightly higher temperature and high abundance regime that is comparable to the measurements in N1. The abundances then decrease to a regime where a majority of the hot cores lie, with N5, most of the W51 hot cores, \deepsouth, and WB 89789 SMM1 occupying overlapping areas of the parameter space. Sgr B2(N4) at $190$ K and W51 e2 at $320$ K are outliers, with factors of several lower and higher abundances when compared to sources in nearby temperature bins, respectively.}
}
{\citet{Bonfand2019} model the time-dependent chemical evolution of the hot cores Sgr B2(N2-N5) in order to constrain the physical conditions that led to the chemical abundances observed in those cores. They found that relatively high dust temperatures ($\geq25$ K) and a slightly lower cosmic-ray ionization rate than predicted for the CMZ is needed to reproduce the observed abundances of \methanol and other COMs in their work. Given the similar peak abundance and coincident temperature in several DS hot cores, this may suggest similar physical conditions were present during their formation. The conditions responsible for the abundances we observe at higher temperatures ($\gtrsim250$ K), conversely, are less clear.  Per \citet{Garrod+2022}, reactions with H$_3$O$^+$ and other species are predicted to decrease \methanol abundance at temperatures beyond the desorption of water ice ($\gtrsim200$ K). Instead, we observe a roughly constant distribution of abundances (to within factors of a few) across DS and a peak abundance that is nearly an order of magnitude higher than DS in W51 e2. This suggests that there are high-temperature formation pathways for \methanol, or that its destruction mechanisms are inhibited at these temperatures. Time-dependent modeling of the full chemical composition of the DS hot cores and the other cores in this comparison set would likely shed light on these questions, as would a larger sample of hot cores with which to compare.}

\section{Conclusions} \label{sec:conclusion}
In this work, we report the discovery and resolved physical properties of nine new hot molecular cores throughout \deepsouth and the hot core \sgrbtwosouth. We summarize our main results here: 
\begin{itemize}
    \item {These measurements reveal sources that have structural properties that are consistent with hot cores in the Galactic disk, with comparable radii and indications of similar interior structure below our resolution limit. However, our sample is also more massive and somewhat warmer than hot cores with similar measurements in the Galactic Disk, and has an average shallower distribution of radial density power law indices. This radial density power-law result may be evidence of the effects of the higher surface density threshold for star formation in the CMZ.}
    \item {\methanol abundances generally decrease with distance from the hot cores and show a positive correlation with temperature across the full sample, consistent with the canonical picture of protostellar thermal feedback sublimating COMs in their envelopes.} %We disfavor \methanol and dust opacity biasing this result due to low optical depth measurements. 
    {Comparing peak abundance in DS to those other other regions in Sgr B2 and the Galaxy, we find a relationship that qualitatively agrees with recent chemical modeling of massive protostars. However, we observe a high-temperature level-off in peak abundance, which suggests there are additional mechanisms of producing or preserving methanol which have yet to be explored.}
    \item We find that a subset of Deep South hot cores shares physical characteristics (e.g. Class II \methanol masers, UCHII regions) with hot cores found in Sgr B2(N), suggesting these cores may be at similar evolutionary stages. Three of the discovered hot cores are associated with Class II \methanol masers, and two are associated with 4.83 GHz H$_2$CO masers. %These sources may be useful in studying the relationship between evolutionary state and the presence of outflows and different types of masers.
    \item Five of our newly discovered hot cores and \sgrbtwosouth exist in a radius of 1 pc and include a total H$_2$ mass of at least {$8.9\times10^3$ M$_\odot$.} Taken with the many lower-mass YSOs in this radius, our results suggest that \sgrbtwosouth and nearby hot cores may be a less evolved but comparably massive cluster to Sgr B2(N) and (M). This makes (DS) a region of interest for studying the early phases of massive cluster formation.
\end{itemize}

\section{Acknowledgements}
{We thank the anonymous referee for their constructive comments and suggestions which greatly improved the quality of this work.} This paper makes use of the following ALMA data: ADS/JAO.ALMA\#2017.1.00114.S. ALMA is a partnership of ESO (representing its member states), NSF (USA) and NINS (Japan), together with NRC (Canada), MOST and ASIAA (Taiwan), and KASI (Republic of Korea), in cooperation with the Republic of Chile. The Joint ALMA Observatory is operated by ESO, AUI/NRAO and NAOJ. The National Radio Astronomy Observatory is a facility of the National Science Foundation operated under cooperative agreement by Associated Universities, Inc. Support for this work was provided by the NSF through the Grote Reber Fellowship Program administered by Associated Universities, Inc./National Radio Astronomy Observatory. A.G. and D.J. gratefully acknowledge support from NSF grant 2008101 and from the NRAO under the SOS program.
A.G. acknowledges support from the NSF under grants 2142300 and 2206511. M.B. is a postdoctoral fellow in the University of Virginia’s VICO collaboration and is funded by grants from the NASA Astrophysics Theory Program (grant number 80NSSC18K0558) and the NSF Astronomy \& Astrophysics program (grant number 2206516).
C.B.  gratefully  acknowledges  funding  from  National  Science  Foundation  under  Award  Nos. 1816715, 2108938, 2206510, and CAREER 2145689, as well as from the National Aeronautics and Space Administration through the Astrophysics Data Analysis Program under Award No. 21-ADAP21-0179 and through the SOFIA archival research program under Award No.  09$\_$0540.
A.S.-M. acknowledges support from the RyC2021-032892-I grant funded by MCIN/AEI/10.13039/501100011033 and by the European Union `Next GenerationEU’/PRTR, as well as the program Unidad de Excelencia María de Maeztu CEX2020-001058-M, and support from the PID2020-117710GB-I00 (MCI-AEI-FEDER, UE).

\bibliographystyle{aasjournal}

\appendix

\section{Integrated Intensity Maps}\label{app:moment0s}
In Figure 16, we show integrated intensity maps for each of the \methanol transitions that were incorporated into the rotational diagrams, as reported in Section \ref{subsec:trotmeasurement}. Figure 16.1 shows the transitions used for \sgrbtwosouth, and the following Figures 16.2 to 16.10 show \dsi through \dsix. The complete figure set (10 images) is available in the online journal. Red contours show the 1 mm continuum for each hot core at 3, 6, 8, 12, and 32$\sigma$. 
\begin{figure*}[h]
    \centering
    \includegraphics[scale=0.5,clip=True,trim={2cm 8cm 1cm 8cm}]{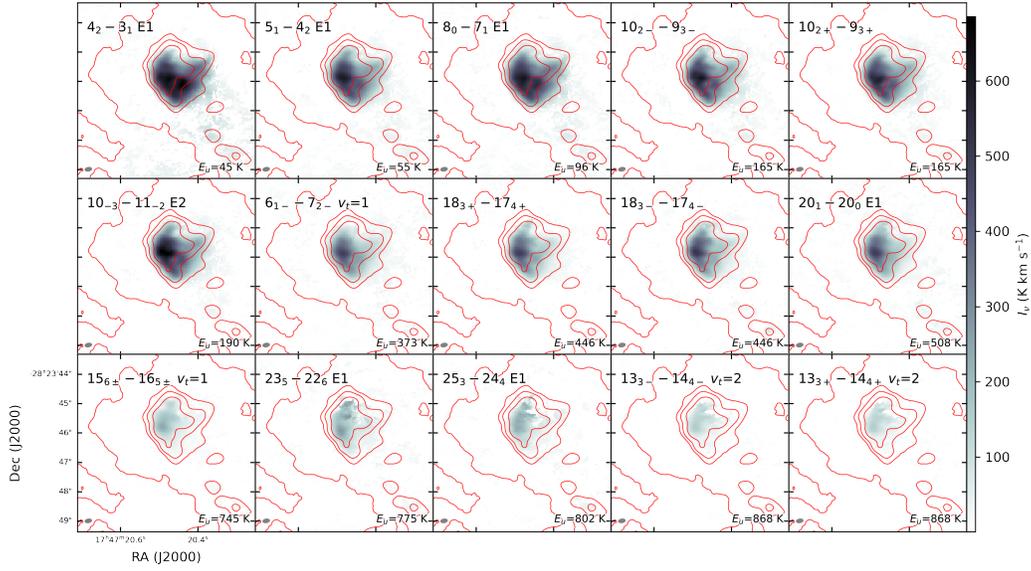}
    \caption{Integrated intensity maps of the detected \methanol emission lines in \sgrbtwosouth. E$_u$ increases from left to right.}
    \label{fig:sgrb2smom0}
\end{figure*}

\clearpage
\section{$n_{transition}$ Maps}\label{app:ntransition}
Figure \ref{fig:dsiiintrans} includes the $n_{transition}$ maps for the sources not shown in Figure \ref{fig:transitionmaps}. The maps are arranged with the left column showing sources \dsii, \dsiv, \dsvi, and \dsviii, and the right column showing sources \dsiii, \dsv, \dsvii, and \dsix. 
\begin{figure}[!h]
    \centering
    \setcounter{figure}{16}
    \includegraphics[scale=0.45]{f27.pdf}
    \includegraphics[scale=0.45]{f28.pdf}
    \includegraphics[scale=0.45]{f29.pdf}
    \includegraphics[scale=0.45]{f30.pdf}
    \includegraphics[scale=0.45]{f31.pdf}
    \includegraphics[scale=0.45]{f32.pdf}
    \includegraphics[scale=0.45]{f33.pdf}
    \includegraphics[scale=0.45]{f34.pdf}
    \caption{\methanol $n_{transition}$ maps for {\dsi and \dsiii} through \dsix, as in Figure \ref{fig:transitionmaps}.}
    \label{fig:dsiiintrans}
\end{figure}
\clearpage
\section{Rotational Temperature Maps}\label{app:texmaps}
In Figure 18, we show the rotational temperature maps for each of the sources not discussed in Figure \ref{fig:texmaps}. The left column, from top to bottom, shows sources \dsii, \dsiv, \dsvi, and \dsviii. The right column shows sources \dsiii, \dsv, \dsvii, and \dsix. 
\begin{figure}[!h]
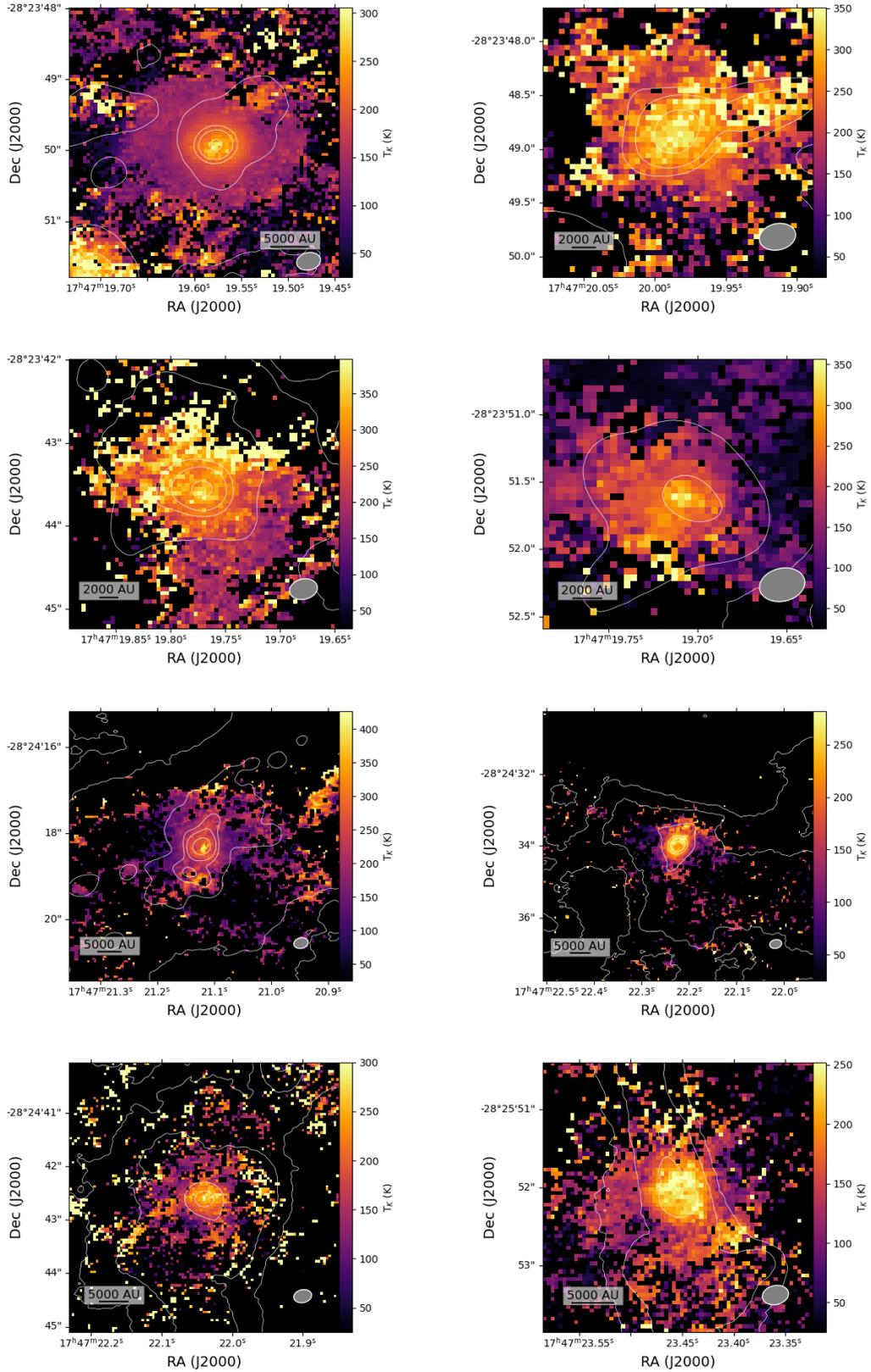

    \centering
    \includegraphics[scale=0.45]{f35.pdf}
    \includegraphics[scale=0.45]{f36.pdf}
    \includegraphics[scale=0.45]{f37.pdf}
    \includegraphics[scale=0.45]{f38.pdf}
    \includegraphics[scale=0.45]{f39.pdf}
    \includegraphics[scale=0.45]{f40.pdf}
    \includegraphics[scale=0.45]{f41.pdf}
    \includegraphics[scale=0.45]{f42.pdf}
    \caption{$T_{rot}$ map for \dsi and \dsiii through \dsix, as in Figure \ref{fig:texmaps}. White contours show the 1 mm continuum at 3$\sigma$, 6$\sigma$, 8$\sigma$, 16$\sigma$, and 32$\sigma$ levels ($\sigma$ = 0.2
mJy beam$^{-1}$).}
    \label{appfig:trotmaps}
 \end{figure}
\clearpage
\section{Methanol Line Panel Plots}\label{app:multilineplots}
Figure 19 shows the \methanol lines used to make the temperature maps for the sources not discussed in Figure \ref{fig:multilineplot} (\dsi-\dsix), with each set of spectra pulled from the representative pixel of the source. The complete figure set (10 images) is available in the online journal.
\begin{figure*}[!h]
    \centering
    \includegraphics[scale=0.5]{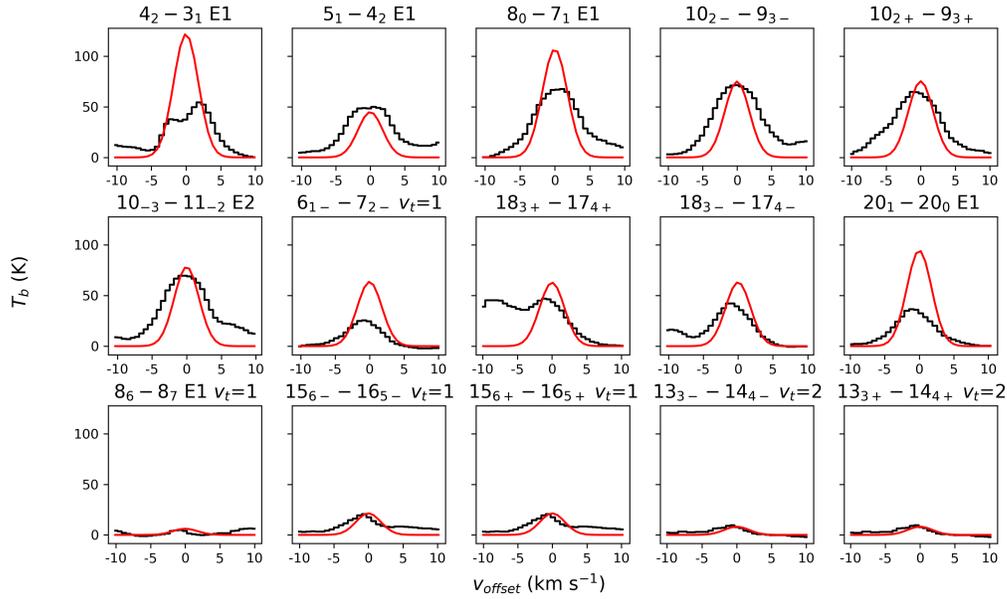}
    \caption{{The same as Figure \ref{fig:multilineplot} for \dsi.}}
    %\label{fig:multilineplot}
\end{figure*}

\clearpage
\section{Spectra}\label{app:spectra}
{Figure 20 shows spectra from each source's representative pixel (see Table \ref{table:obsprops}), as discussed in Section \ref{sec:temps}. The spectra are pulled from all four spectral windows in the data set and arranged in pairs, with spw0 and spw1 for a given core shown first and spw2 and spw3 shown in the following figure. The complete figure set (20 images, Figures 20.1-20.20) is available in the online journal.}
    %\centering
    %\vspace*{10cm}
    %\includegraphics[width=\textwidth]{f44.pdf}\\
    %\vspace*{-2cm}
    %\includegraphics[width=\textwidth]{f45.pdf}%, inner]
    %\caption{Example spectra from the hot cores \sgrbtwosouth and \dsii showing the richness of their chemical composition. The black lines are the 1 mm ALMA spectral data. Colored lines are model spectra for \methanol and a selection of tentatively identified species created using the \methanol-derived $T_{rot}$ measurement in the target pixel, by-eye estimates of $N_{tot}$, and the \texttt{lte\_molecule} package of \texttt{pyspeckit}. \methanol lines are marked by the dashed green model lines.}
    %\label{fig:examplespec}

\begin{sidewaysfigure*}[!h]
    \centering
    \includegraphics[width=\textwidth]{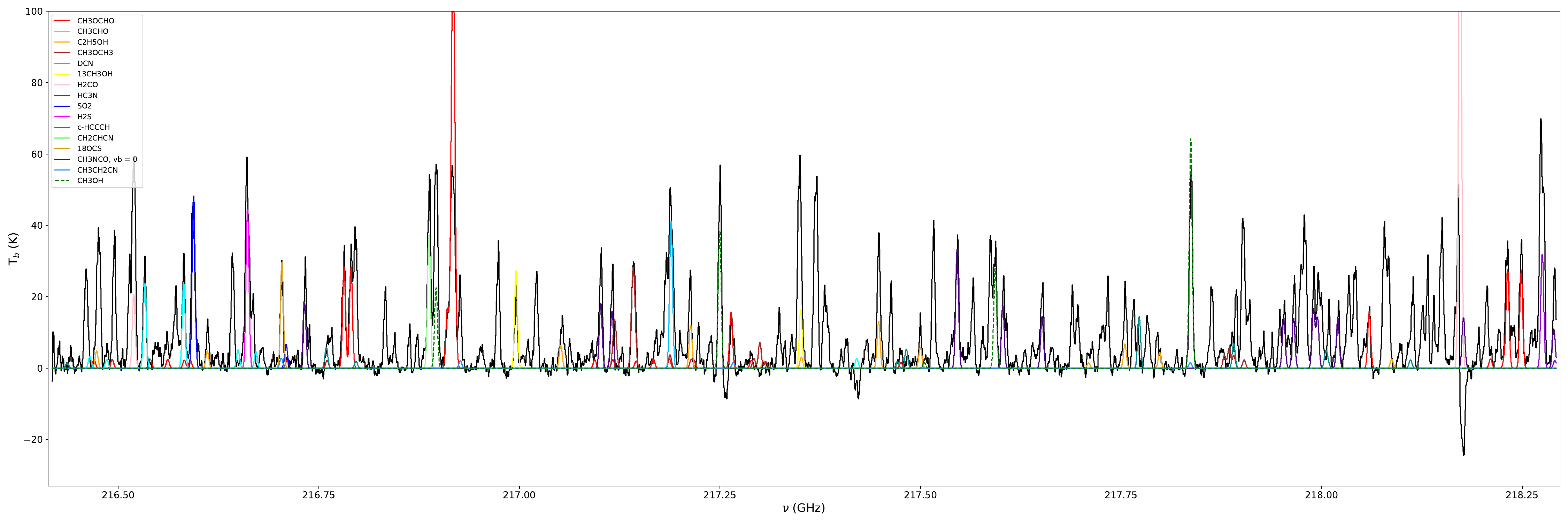}\\
    %\vspace*{-2cm}
    \includegraphics[width=\textwidth]{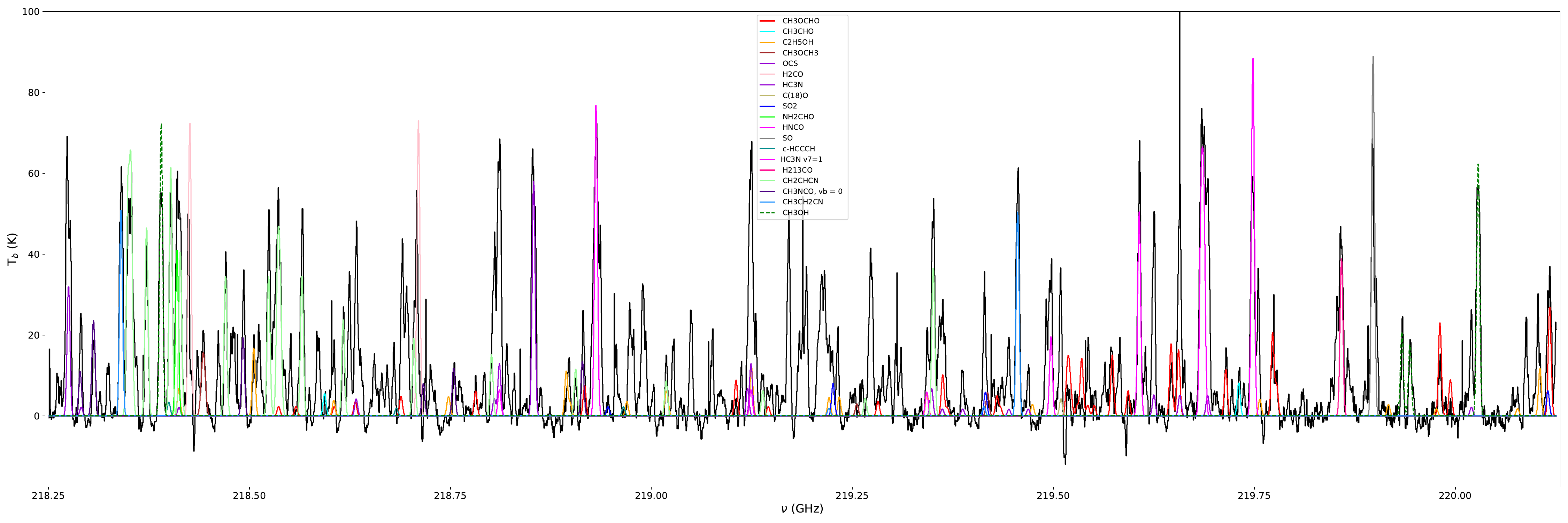}%, inner]
    \caption{Spectra from the representative pixel of \sgrbtwosouth showing the richness of its chemical composition. The black lines are the 1 mm ALMA spectral data. Colored lines are model spectra for \methanol and a selection of tentatively identified species created using the \methanol-derived $T_{rot}$ measurement in the target pixel, by-eye estimates of $N_{tot}$, and the \texttt{lte\_molecule} package of \texttt{pyspeckit}. \methanol lines are marked by the dashed green model lines.} \label{fig:examplespec}%\label{appfig:sgrb2southspec01}
\end{sidewaysfigure*}
\clearpage
\section{Radial Temperature Profiles}\label{appendix:texprofs}
Figure 21 shows the radial temperature profiles for each source not discussed in Figure \ref{fig:radtempprof}. The profiles are arranged in the same order as in Figure 17, with the left column showing sources \dsii, \dsiv, \dsvi, and \dsviii, and the right column showing sources \dsiii, \dsv, \dsvii, and \dsix.
\begin{figure}[!h]
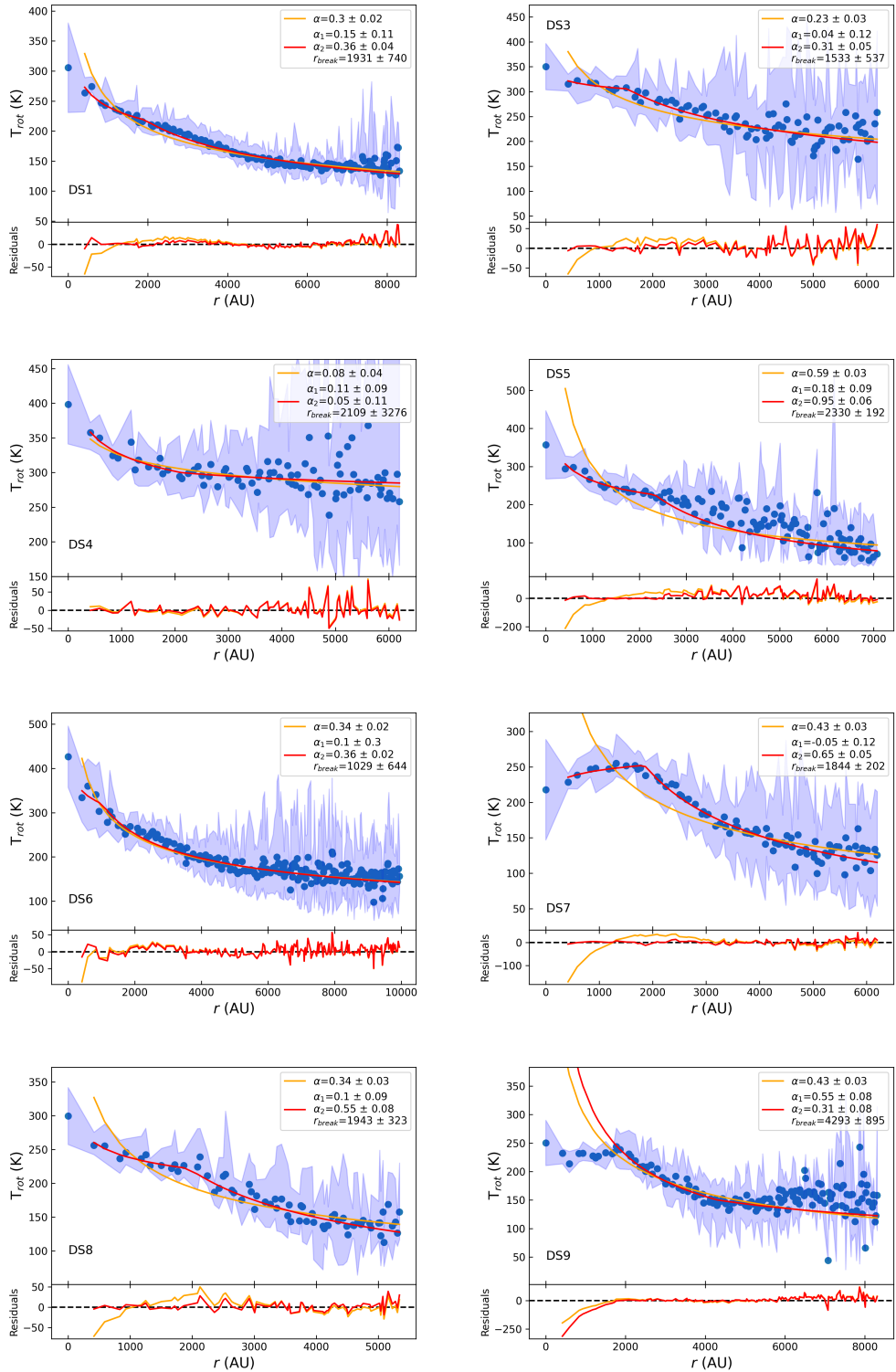

    \centering
    \setcounter{figure}{20}
    \includegraphics[scale=0.42]{f48.pdf}
    \includegraphics[scale=0.42]{f49.pdf}
    \includegraphics[scale=0.42]{f50.pdf}
    \includegraphics[scale=0.42]{f51.pdf}
    \includegraphics[scale=0.42]{f52.pdf}
    \includegraphics[scale=0.42]{f53.pdf}
    \includegraphics[scale=0.42]{f54.pdf}
    \includegraphics[scale=0.42]{f55.pdf}
    \caption{The same as Figure \ref{fig:radtempprof} but for {\dsi and \dsiii} through \dsix.}
    \label{fig:appendixradtex}
\end{figure}%(Upper) Azimuthally-averaged temperature distribution for \dsii. The blue shaded region represents the range of temperatures observed in each radial bin. Single and broken power law profiles are plotted in orange and red, respectively. (Lower) The residuals of the two models. The broken power law model has lower residuals throughout the distribution

\clearpage
\section{$N_{H_2}$ Maps}\label{app:nh2maps}
Figure 22 contains $N_{H_2}$ maps for the hot cores in our sample, as discussed in Section \ref{subsec:radialdensityprofiles}. Contours show the 1 mm continuum for each hot core with levels at 3, 6, 8, 12, and 32$\sigma$.
\begin{figure}[!h]
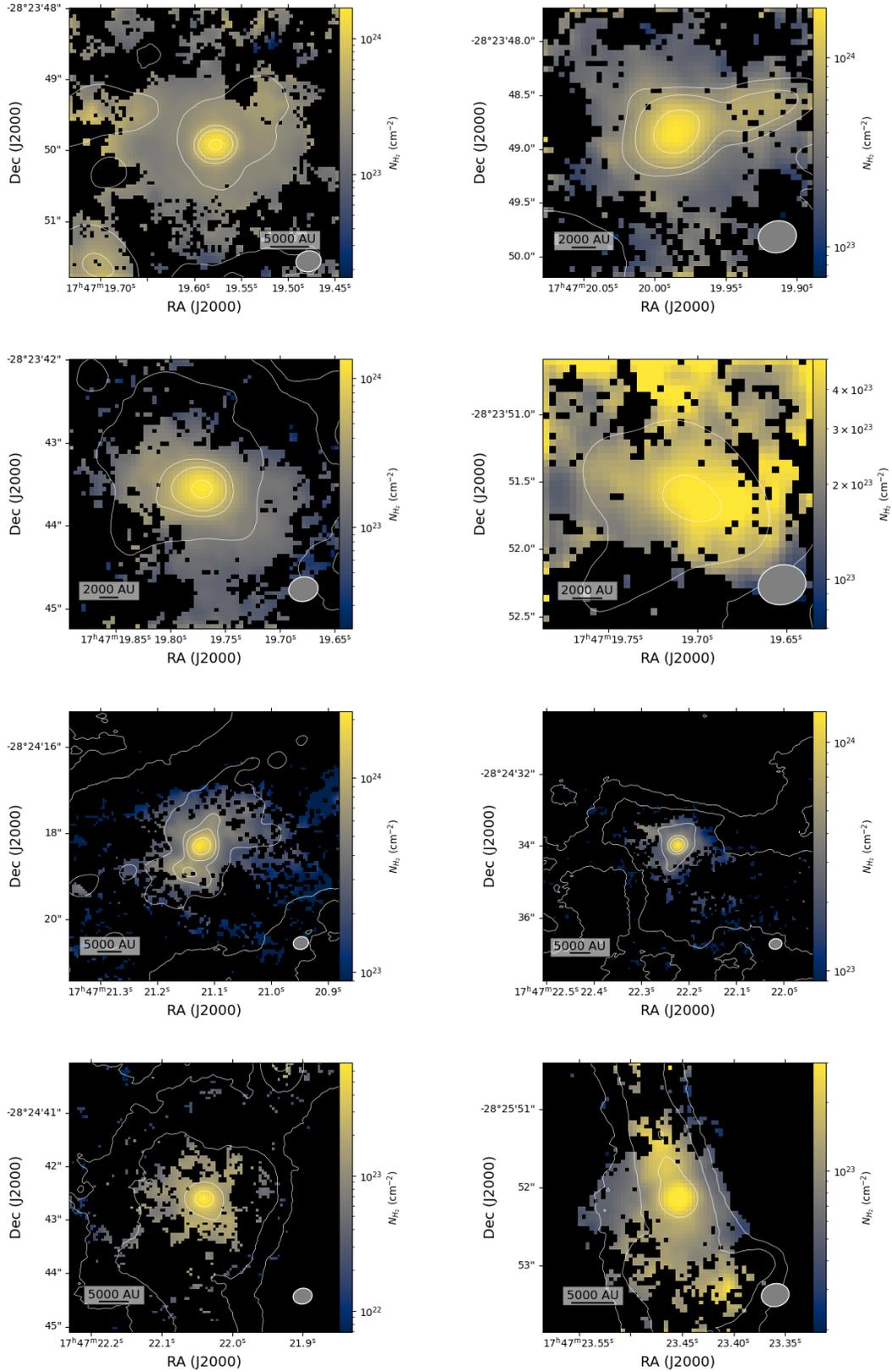

    \centering
    \includegraphics[scale=0.45]{f56.pdf}
    \includegraphics[scale=0.45]{f57.pdf}
    \includegraphics[scale=0.45]{f58.pdf}
    \includegraphics[scale=0.45]{f59.pdf}
    \includegraphics[scale=0.45]{f60.pdf}
    \includegraphics[scale=0.45]{f61.pdf}
    \includegraphics[scale=0.45]{f62.pdf}
    \includegraphics[scale=0.45]{f63.pdf}
    \caption{$N_{H_2}$ maps for {\dsi and \dsiii} through \dsix. White contours show the 1 mm continuum at 3$\sigma$, 6$\sigma$, 8$\sigma$, 16$\sigma$, and 32$\sigma$ levels ($\sigma$ = 0.2
mJy beam$^{-1}$).}
    \label{appfig:trotmaps}
 \end{figure}
\clearpage
\section{Abundance Maps}\label{app:abundmaps}
Figure 23 contains X(\methanol) maps for the sources not shown in Figure \ref{fig:abundancemaps}. The maps are arranged in the same order as in Figure \ref{appfig:trotmaps}, with the left column showing sources \dsii, \dsiv, \dsvi, and \dsviii, and the right column showing sources \dsiii, \dsv, \dsvii, and \dsix. 
\begin{figure}[!h]
    \centering
    \includegraphics[scale=0.45]{f64.pdf}
    \includegraphics[scale=0.45]{f65.pdf}
    \includegraphics[scale=0.45]{f66.pdf}
    \includegraphics[scale=0.45]{f67.pdf}
    \includegraphics[scale=0.45]{f68.pdf}
    \includegraphics[scale=0.45]{f69.pdf}
    \includegraphics[scale=0.45]{f70.pdf}
    \includegraphics[scale=0.45]{f71.pdf}
    \caption{X(\methanol) map for {\dsi and \dsiii} through \dsix, as in Figure \ref{fig:abundancemaps}.}
    \label{appfig:abundancemaps}
\end{figure}
\clearpage
\section{Abundance versus Temperature Distributions}\label{app:abundvstex}
Figure 24 includes the X(\methanol) vs $T_{rot}$ distributions for the sources not shown in Figure \ref{fig:abundancevstemperature}. The left column includes sources \dsi, \dsiv, \dsvi, and \dsviii. The right column includes sources \dsiii, \dsv, \dsvii, and \dsix.
\begin{figure}[!h]
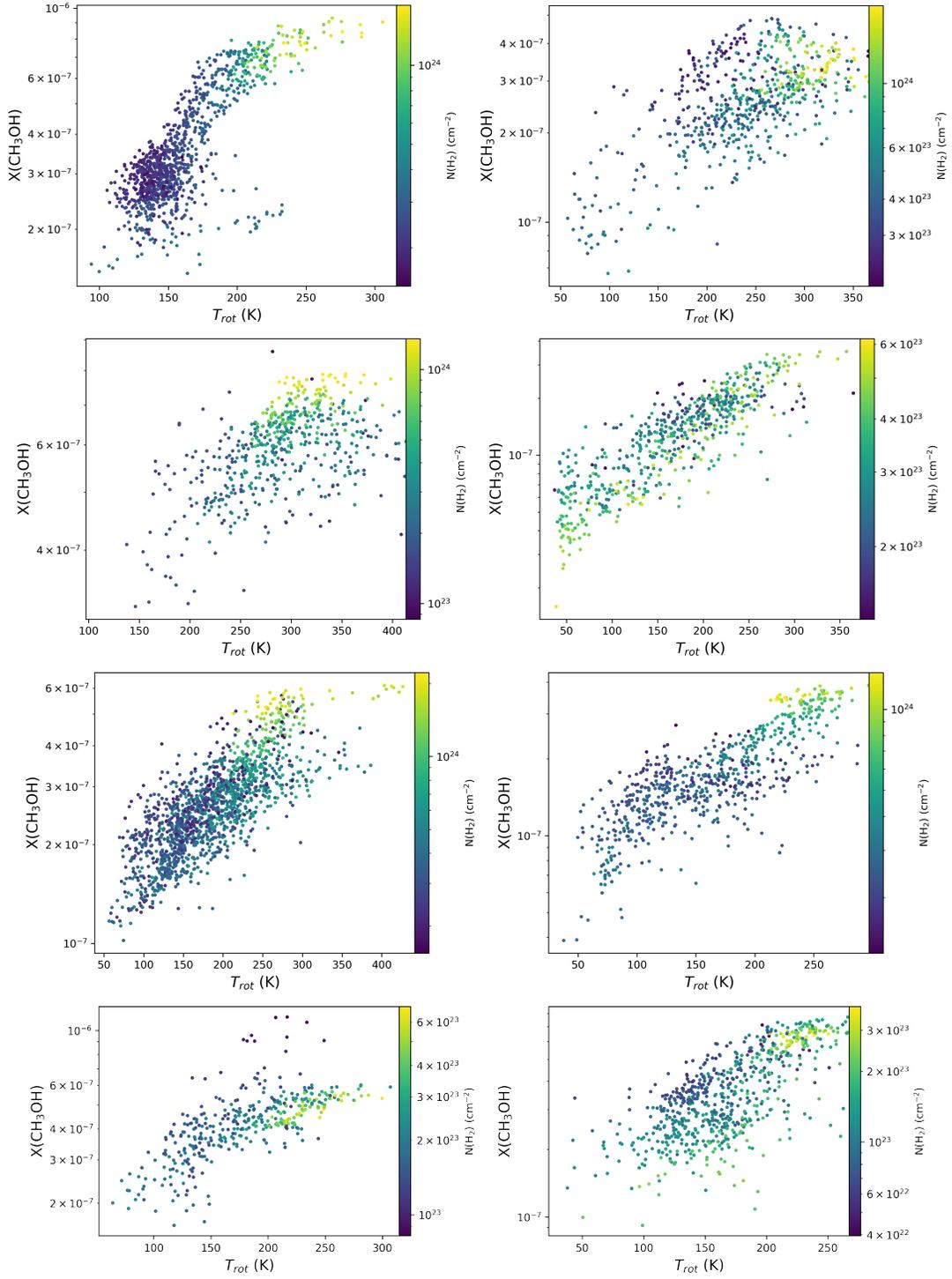

    \centering
    \includegraphics[scale=0.45]{f72.pdf}
    \includegraphics[scale=0.45]{f73.pdf}
    \includegraphics[scale=0.45]{f74.pdf}
    \includegraphics[scale=0.45]{f75.pdf}
    \includegraphics[scale=0.45]{f76.pdf}
    \includegraphics[scale=0.45]{f77.pdf}
    \includegraphics[scale=0.45]{f78.pdf}
    \includegraphics[scale=0.45]{f79.pdf}
    \caption{The same as Figure \ref{fig:abundancevstemperature} for \dsi and {\dsiii} through \dsix.}
    \label{fig:dsivtexabun}
\end{figure}

\end{document}